# Incorporation of magnetic nanoparticles into lamellar polystyrene-b-poly(n-butyl methacrylate) diblock copolymer films: influence of the chain end-groups on nanostructuration

Siham Douadi-Masrouki[a, b, c, 1], Bruno Frka-Petesic [a, b, c, 2], Maud Save[d, e, 3], Bernadette Charleux[d, e, 4], Valérie Cabuil[a, b, c, 5], Olivier Sandre*,[a, b, c, 6]

[a] UPMC Univ Paris 6, UMR 7195 Physicochimie des Electrolytes, Colloïdes et Sciences Analytiques – 4 place Jussieu, case 51 75005 Paris France

[b] Centre National de la Recherche Scientifique, UMR 7195 PECSA, 75005 Paris France

[c] ESPCI ParisTech, UMR 7195 PECSA, 75005 Paris France

[d] UPMC Univ Paris 6, UMR 7610 Laboratoire de Chimie des Polymères – 4 place Jussieu, case 185 75005 Paris France

[e] Centre National de la Recherche Scientifique, UMR 7610 LCP, 75005 Paris France

*corresponding author: olivier.sandre@enscbp.fr, tel: +33-5-4000-3695, fax: +33-5-4000-8487

**Abstract:** In this article, we present new samples of lamellar magnetic nanocomposites based on the self-assembly of a polystyrene-*b*-poly(n-butyl methacrylate) diblock copolymer synthesized by atom transfer radical polymerization. The polymer films were loaded with magnetic iron oxide nanoparticles covered with polystyrene chains grown by surface initiated-ATRP. The nanostructuration of the pure and magnetically loaded copolymer films on silicon was studied by atomic force microscopy, ellipsometry, neutron reflectivity and contact angle measurement. The present study highlights the strong influence of the copolymer extremity – driven itself by the choice of the ATRP chemical route – on the order of the repetition sequences of the blocks in the multi-lamellar films. In addition, a narrower distribution of the nanoparticles' sizes was examined as a control parameter of the SI-ATRP reaction.

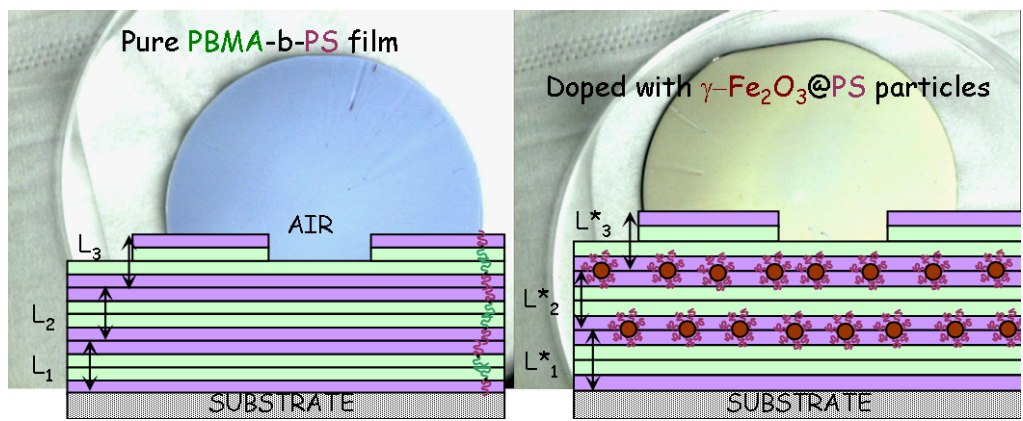

**Graphical Abstract:** Sketches of the nanostructure of lamellar films of PBMA425-*b*-PS490 block copolymer either pure or doped with magnetic nanoparticles coated by a PS brush and macroscopic pictures of the films on silicon wafers (respectively nb. 2 and 13 in Table 6).

---

[1] siham.douadi@gmail.com

[2] bruno.frka-petesic@gmail.com

[3] Present address: Université de Pau et des Pays de l'Adour – CNRS, UMR 5254 IPREM Equipe Physique et Chimie des Polymères – Technopole Hélioparc, 2 Av P. Angot, 64053 Pau France, maud.save@univ-pau.fr

[4] Present address: Université Claude Bernard Lyon 1 – CNRS – CPE, UMR 5265 Chimie, Catalyse, Polymères et Procédés – CPE Lyon Bât 308, 43 Bd du 11 Novembre 1918, 69616 Villeurbanne France, bernadette.charleux@lcpp.cpe.fr

[5] valerie.cabuil@upmc.fr

[6] Present address: Univ. Bordeaux – CNRS – Bordeaux INP, UMR 5629 Laboratoire de Chimie des Polymères Organiques – ENSCBP 16 Av Pey Berland, 33600 Pessac France, olivier.sandre@enscbp.fr

*1. Introduction*

In order to design new materials with multi-functional properties, chemists elaborate composite materials combining an organic matrix with inorganic fillers. While being homogeneous at the macroscopic scale, those materials usually contain characteristic internal structures of different geometries (points, cylinders, planar or curved surfaces… ) that can exhibit a long-range order (*e.g.* lamellar or cylindrical mesophases) responsible for the improvement of physical properties (mechanical resistance, special anisotropy, response to specific stimuli…). One elegant way to engineer a nanostructured matrix loaded with inorganic nanoparticles somehow organized at a large scale relies on their co-assembly with block copolymers (BCP) into complex architectures.[1] Our work focus on the case of lamellar materials where the alternating layers are made of diblock copolymers and iron oxide magnetic nanoparticles (MNP), which open new applications as optical coatings with specific reflection or guiding properties. An ordered polymer matrix with a lamellar morphology at the mesoscopic scale can be obtained by depositing a melt of a symmetrical diblock copolymer onto a flat substrate. In spite of the weak incompatibility of polystyrene (PS) and poly(n-butyl methacrylate) (PBMA), the existence of ordered layers has been shown for molten films of symmetrical PS-*b*-PBMA diblock copolymers of different molar masses provided that they are annealed at an appropriate temperature.[2] According to the literature on thin nanocomposite films, one can maintain a lamellar morphology while doping certain blocks of the BCP if the volume fraction of inorganic nanoparticles is sufficiently low and if their diameters are smaller than a fraction of the pure BCP lamella thickness (typically 20-30%).[3] In addition, the surface of the nanoparticles must be grafted by a polymer of the same nature as one of the blocks, as it was shown for gold nanoparticles with various BCPs.[4] When using MNPs, such hybrid lamellar materials are of particular interest for technological applications, because they combine the low processing temperature of the polymer matrix (compared to standards in electronics industry) with the outstanding properties of magnetic fillers. On the one hand, MNPs interact strongly with electromagnetic waves over a broad spectrum of wavelengths: magnetic birefringence, circular dichroism and Faraday rotation in the UV-Vis range,[5] Brown's and Néel's relaxations in radiofrequency (kHz-MHz) magnetic fields inducing hyperthermia,[6] ferromagnetic resonance (FMR) causing a strong absorption of microwaves (GHz).[7] On the other hand, a long-range order as multi-layers of MNPs could select propagation modes of the waves and give rise to Bragg-type reflection also called "optical-band gaps" or "photonic crystals".[8] In this article, we focus our attention on the insertion into a lamellar symmetrical BCP film of iron oxide MNPs, exhibiting superparamagnetism. Those MNPs behave indeed as "soft magnets", each of them bearing a large magnetic moment (typically $10^4$ Bohr's magnetons) that orients in an external magnetic field but retains no magnetization in a zero field due to thermal agitation.

This work differs from reported studies that dealt with metallic MNPs (FePt alloy) deposited at the surface of thin Polystyrene-b-poly(methyl methacrylate) films:[9] on the one hand the PS-*b*-PMMA lamellae were not parallel but perpendicular to the substrate, on the other hand the FePt MNPs were not confined by a chemical coating but rather trapped physically by corrugations of the surface. The goal of those studies was indeed to take benefit from a lamellar period of 30 nm to control the surface density of thermally blocked magnetic dipoles of the FePt face-centered tetragonal phase[10] (above 500°C) in order to get magnetic storage materials with increased capacity. Other early studies on thin films of symmetrical polystyrene-*b*-poly(n-butyl methacrylate) copolymers have underlined the importance of the conditions of film deposition (*e.g.* substrate smoothness) and annealing (temperature, hydrostatic pressure and time length) on the quality of the lamellar structure. Their authors examined the defects at the top surface of the films[11, 12], the width and height of the scattering peak[13] and the lateral correlation length ξ of the nanostructure.[14] Later studies described lamellar magnetic nanocomposites made from PS-*b*-PBMA and iron oxide MNPs synthesized separately and then co-assembled by depositing a mixture in solution onto a flat substrate and annealing the films above the Tg of the PS blocks.[15-18] Other teams studied the alternative route that consists in using a preformed lamellar structure of BCP as the template for the synthesis of MNPs either from a vapour[19] or a liquid phase[20], as it was also proposed for an isotropic PS matrix[21] or for poly(acrylic acid)-*g*-poly(n-butyl acrylate) cylindrical brushes[22]. The studies on PS-*b*-PBMA were continued for doped films by the quantification of the distortion by the MNPs of the lamellar order (through the decrease of ξ)[23] and by the kinetics of growth of defects at the film surface for undoped ones.[24] Until now, all the articles on nanostructuration with PS-*b*-PBMA share the common feature of utilizing copolymers synthesized by anionic polymerization. The most recent article on PS-*b*-PBMA revealed the strong influence of the chemical nature of the chain end-groups on its phase diagram.[25] More precisely, maleic anhydride end-groups induce the perforation of the lamellae above 245°C, which is not observed for carboxylate end-groups. An important question raised by our present work is the influence of end-groups resulting from the polymerization scheme (controlled radical *vs.* anionic polymerization) onto the wetting properties of the lamellae both on the hydrophilic substrate and at the top layer. Whereas the synthesis of PS-*b*-PBMA diblock copolymers by anionic polymerization is well documented,[26-28] only one article described the synthesis of PS-*b*-PBMA diblock copolymers by controlled radical polymerization (CRP).[29] Among the CRP methods, atom transfer radical polymerization (ATRP) is a convenient pathway to synthesise well-defined methacrylic based copolymers such as PMMA-*b*-PS 30 or P*i*-BMA-*b*-PS.[31] We report here on the synthesis of well-defined PBMA-*b*-PS symmetric diblock copolymers synthesized by ATRP in bulk. The challenge consisted in synthesizing well-defined PBMA-

*b*-PS diblock copolymers *via* ATRP with sufficient high molar masses to reach the segregated state. In most of the previous studies, the MNPs made of maghemite iron oxide were coated with a PS brush of molar mass 13000 g.mol-1 obtained by a "grafting onto" technique[32] using sulfonate-terminated PS chains also synthesized by anionic polymerization and end-functionalized by 1,3-propane sultone.[33] Here we describe an efficient synthesis of the same type of inorganic magnetic nanoparticles grafted with a PS shell by surface initiated polymerization also called "grafting from". Due to the popular success of ATRP as a versatile controlled radical polymerization technique, the coating of iron oxide nanoparticles with different polymers using surface initiated (SI)-ATRP has been described recently by other groups.[34-36] The add-on value of our study consists in polymerizing PS from the surface of MNPs with only moderate aggregation issue and inserting them into the lamellae of PBMA-*b*-PS while keeping the long-range order as evidenced by several surface techniques such as ellipsometry, contact angle measurement and neutron reflectivity. The synthesis of PBMA-*b*-PS by ATRP offers us the opportunity to study the influence of a halide chain end-group onto the lamellar ordering of the diblock copolymer, in comparison with the previous studies[11-18, 23-27, 29] where the PS-*b*-PBMA samples were synthesized by anionic polymerization, leading to H-terminated chains[27] (see Scheme 1 in the Electronic Supplementary Material file).

## 2. Experimental

### 2.1 Materials

Styrene (S) and *n*-butyl methacrylate (BMA) (Aldrich) were distilled prior to use in order to remove the inhibitor. *N,N′,N″,N″*-pentamethyldiethylenetriamine (PMDETA), copper (I) bromide CuBr, copper (II) bromide CuBr2, copper (I) chloride CuCl, copper (II) chloride CuCl2 (Aldrich), benzonitrile, ethyl-2-bromoisobutyrate (EBrIB) (Aldrich), hexane and toluene were used as received.

### 2.2 Synthesis of PBMA-b-PS diblock copolymer

*Synthesis of PBMA macroinitiator by ATRP.* A solution of BMA *(15 g, 10.5 mmol),* PMDETA *(46 mg, 0.26 mmol)*, and EBrIB *(51 mg, 0.26 mmol)* was degassed by nitrogen bubbling during 30 minutes. The degassed solution was transferred under a nitrogen flow *via* a double-tipped stainless steel needle into the polymerization flask containing the copper halide. The reactor was immediately immersed in an oil bath thermostated at 100°C for the required polymerization time. The polymer solution was diluted in dichloromethane and passed through a neutral alumina column to remove the copper catalyst. The recovered PBMA homopolymer solution was precipitated in a large excess of methanol, and dried under vacuum at room temperature.

*Synthesis of the PBMA-b-PS diblock copolymer by ATRP.* A solution containing the PBMA macroinitiator *(5 g, 0.11 mmol),* styrene *(11,13 g, 10.7 mmol)* and PMDETA *(19 mg, 0.11 mmol)* was prepared prior to be degassed by nitrogen bubbling during 30 minutes. The introduction of the copper halide (CuCl) under nitrogen flow into the polymerization flask thermostated at 100°C marked the time zero of the polymerization. After the required time of reaction, the crude polymer solution diluted with dichloromethane was passed through a neutral alumina column. The recovered diblock copolymer was subsequently precipitated into methanol and the residual PBMA block was removed by extraction with hexane.

### 2.3 Synthesis of γ-Fe$_2$O$_3$@PS nanoparticles

*Aqueous synthesis of γ-Fe$_2$O$_3$ nanoparticles.*
 Superparamagnetic nanoparticles made of maghemite (γ-Fe$_2$O$_3$) were synthesized in water according to Massart's procedure.[37] At first, magnetite Fe$_3$O$_4$ nanocrystals (also called ferrous ferrite FeO.Fe$_2$O$_3$) were prepared from an alkaline coprecipitation of a quasi-stoichiometric mixture of iron +II (1 mole) and iron +III (1.4 mole) salts in an acidic medium (HCl pH≈0.4). One litre of a concentrated ammonia solution (20 %) was quickly added onto the acidic iron salts mixture, which produced a black solid suspension almost instantaneously. Those Fe$_3$O$_4$ nanoparticles were first acidified with 0.36 L of nitric acid (52%) then oxidized by adding a ferric nitrate solution (0.8 mole). After 30 min at boiling temperature, the suspension had turned to a red colour characteristic of maghemite γ-Fe$_2$O$_3$ nanoparticles. After washing steps in acetone and diethyl-ether to remove the excess ions, the nanoparticles are readily dispersed in water and form a true "ionic ferrofluid", made of maghemite nanoparticles with a positively charged surface, which remain in a monophasic state under the application of a magnetic field of arbitrary value. Those crystals exhibit a Log-Normal distribution of diameters with an average around 7 nm and a standard deviation $\sigma$ about 0.4, as measured by magnetometry (ESI-Figure 1). This aqueous ferrofluid is then coated by oleic acid to obtain an oily ferrofluid,[38] which is a dispersion of magnetic nanoparticles stabilized by steric repulsions in an organic medium (n-hexane) compatible for the synthesis of γ-Fe$_2$O$_3$@PS nanoparticles. Briefly, the grafting reaction was performed by mixing the aqueous MNPs with 20 mol% of ammonium oleate (firstly obtained by deprotonation of oleic acid with ammonia) for 30 min at 70°C. The resulting precipitate was washed 3 times with methanol and finally with diethyl ether in order to remove the excess surfactant and all traces of water until redispersion in n-hexane. Once coated by oleate, the MNPs exhibit a low hydrodynamic diameter in n-hexane (between 26 and 36 nm depending on the samples as measured by DLS in dilute suspensions) with a solid content that can be concentrated up to 30 % w/w, the solid weight being composed at 75±5% of iron oxide as evidenced by iron titration and gravimetry.

In the following experiments, we have used either maghemite nanoparticles with a relatively high size polydispersity ($\sigma$=0.4) or suspensions which have been treated with a size-sorting procedure by fractionated phase-separation enabling to decrease the standard deviation down to $\sigma$=0.2 and that will be named thereafter "monodisperse" magnetic nanoparticles.[39]

*Synthesis of $\gamma$-Fe$_2$O$_3$@PS nanoparticles by surface-initiated polymerization (SI-ATRP).*

As described previously, an ATRP initiator was coated onto the $\gamma$-Fe$_2$O$_3$@oleate nanoparticles *via* a ligand-exchange reaction.[40] More precisely, the surfaced ferrofluid (28.3 mmol iron) was mixed with 2-bromo-2-methylpropionic acid (8.5 mmol BrMPA) in n-hexane (30mL). The atmosphere of the mixture was replaced with nitrogen by freeze-thawing and pumping cycles using a Schlenk tube equipped with a magnetic stirrer and a Rotaflow valve. After 72 h of stirring at room temperature under inert atmosphere, a magnetic precipitate was obtained. After several washes with n-hexane to remove oleic acid, the BrMPA-functionnalized $\gamma$-Fe$_2$O$_3$ nanoparticles were readily dispersed in styrene. To achieve SI-ATRP, the mixture of $\gamma$-Fe$_2$O$_3$@BrMPA, styrene and PMDETA was carefully degassed by nitrogen bubbling during 30 min. Then it was transferred through a cannula into the polymerization flask already containing copper bromide. The reactor was immersed in an oil bath thermostated at 100°C for the required polymerization time. The product was diluted in dichloromethane and filtrated through a Fluoropore™ membrane to remove the copper catalyst[41] or simply washed by water. After this purification step, the $\gamma$-Fe$_2$O$_3$@PS core-shell nanoparticles were precipitated in a large excess of methanol and dried out on a Büchner. The brownish powder could be readily dispersed in dichoromethane or toluene for characterization or further utilization.

Raw and grafted nanoparticles were characterized using vibrating sample magnetometry (VSM), atomic absorption spectroscopy (AAS), transmission electron microscopy (TEM), and dynamic light scattering (DLS) experiments. For analysis purpose, aliquots of each synthesis of $\gamma$-Fe$_2$O$_3$@PS nanoparticles were degraded into iron ions by hot treatment in an acidic mixture of HNO$_3$/HCl that leaves the PS chains intact. Then the polymer was separated by phase transfer into dichloromethane and precipitated in an excess of methanol, in order to be analyzed by size-exclusion chromatography (SEC).

2.4 Analytical techniques

*1H NMR (250 MHz) analyses* were performed in CDCl3 in 5 mm tubes at room temperature using an AC250 Bruker spectrometer. The monomer conversion ($x$) was calculated from the proton NMR spectrum of the crude polymer solution using the vinylic protons of the monomer (5 – 6 ppm) and the characteristic peak of the polymer.

*Size exclusion chromatography (SEC)* was performed in THF at a flow rate of 1 mL.mn-1 to measure the number- ($M_n$) and the weight- ($M_w$) average molar masses and the molar-mass dispersity index ($M_w/M_n$). The SEC apparatus (TDA model 302 from Viscotek) was equipped with two mixed bead columns (PLgel, mixed C, 5µm) and one guard column (PLgel, 5 µm, 100 Å) thermostated at 40°C. Three detectors ran online: a light scattering (LS) detector (2 angles at 90° and 7°, $\lambda$ = 670 nm), a viscosimeter (Wheaston bridge) and a refractive index (RI) detector. SEC with triple detection yields the absolute values of the molar masses without the need of standards, once the refractive index increments (*dn/dc*) of PBMA and PBMA-*b*-PS were measured respectively at 0.077 and 0.125 as the slopes of RI measurements at 40°C *vs.* concentration in THF. Our measured value of *dn/dc* for PBMA is very close to the value 0.075 reported for high molecular weight PBMA synthesized by reverse ATRP in miniemulsion.[42] In the case of PBMA, the most commonly used SEC technique calibrated by polystyrene standards led to underestimated molar masses: $M_n^{triple}/M_n^{standard}$ =1.31±0.01.

*Vibrating sample magnetometry (VSM)* was used to record the magnetization curve $M(H)$ with a home-made apparatus for field values up to 0.9 Tesla, above which a saturation plateau $M$sat was reached. The shape of the normalized curve $M(H)/M_{sat}$ was well fitted by a Langevin's function typical for superparamagnetism, taking into account the size distribution of the nanoparticles.[43] This distribution was well described by a Log-normal function with two parameters: the median diameter $d_0$ and the distribution width $\sigma$. The magnetization at saturation is relied to the volume fraction $\Phi$ by $M$sat= $\Phi$*$m_s$ where $m_s$=3×105 A/m is the specific magnetization of colloidal maghemite, which is lower than for the bulk oxide.[44]

*Atomic Absorption Spectrophotometry (AAS).* The iron concentration was obtained by AAS using a Perkin Elmer AAnalyst 100 apparatus. Before titration, the magnetic nanoparticles ($\gamma$-Fe$_2$O$_3$ or $\gamma$-Fe$_2$O$_3$@PS) were brought up to ebullition in concentrate hydrochloric acid (35%) or a mix of hydrochloric and nitric acids until the total dissolution of the nanoparticles into iron (III) ions. The conversion law from the molar concentration of ions in the titrated aliquot to the volume fraction in the original suspension if simply $\Phi$(v/v %) = 1.577*[Fe] (mol/L) as deduced from the molar mass (159.7 g.mol-1) and the mass density (5.1 g/cm$^3$) of maghemite.

*Transmission electron microscopy (TEM).* The morphology of the particles was determined by TEM using a JEOL 100 CXII (UHR) microscope working at 80kV. To study the $\gamma$-Fe$_2$O$_3$@PS core-shell nanoparticles, we used a short exposition (15 s) of the grids to RuO$_4$ vapour as a contrast agent specific to the aromatic molecules of the PS shell.[45] This agent was produced freshly by reacting 0.2g RuCl$_3$ with 10mL of aqueous NaClO 5.25%.[46]

*Dynamic light scattering (DLS).* Hydrodynamic diameters ($d_h$) were measured with a Malvern NanoZS ZetaSizer working at an angle of 173° in n-hexane (RI $n$=1.375 and viscosity $\eta_{25°C}$=0.28 cP) for the MNPs coated by oleic acid and in dichloromethane (RI $n$=1.432 and viscosity $\eta_{25°C}$=0.389 cP) for the MNPs after grafting by SI-ATRP of PS. The correlograms were fitted using a multi-modal algorithm provided by the manufacturer.

## 2.5 Typical procedure for sample preparation

Silicon wafers (n-doped type from Siltronix Inc., 2" diameter, 355-405 μm thickness) were cleaned as according to the following procedure. At first, silicon substrates were sonicated during 3 minutes in chloroform prior treatment with a freshly prepared "piranha" solution (70/30, v/v, 95% concentrated $H_2SO_4$/ 30% aqueous $H_2O_2$) at 100°C for 10 minutes. Silicon substrates were rinsed with distilled water, ultrapure water and subsequently dried under a nitrogen stream until no trace was visible (otherwise the rinsing step was repeated). The films were deposited by spin-coating 750μL of a toluene solution at 6000 rpm (Süss Microtec Delta-10TT) onto freshly cleaned silicon wafers. The solvent mixtures contained PBMA-*b*-PS and magnetic nanoparticles at a total concentration (if no other specified value) $C$ = 20 g/L with an inorganic volume fraction $\Phi$ (defined as the volume of $Fe_2O_3$ divided by the total dry volume of iron oxide and polymer) between 0 and 0.25%. The films were annealed at 150°C *i.e.* 50°C above the glass temperature of PS, a temperature estimated to be sufficient for self-assembly ($\chi_{150°C}$≈0.04 see 3.1) but 30°C below the order-disorder transition (UCOT) of PBMA-*b*-PS of $M_n$≈$10^5$ g.mol$^{-1}$ according to literature.[2] In addition, the risk of polymer degradation was limited by annealing it under vacuum for a period of 48 or 72 hours at most.

## 2.6 Characterization methods for copolymer films

*Atomic Force Microscopy (AFM).*
The surface of the composite thin films after annealing was characterized by AFM. The images were taken either with a Nanoscope III (Digital Instruments, Veeco) in the dry Tapping Mode™ or with a Picoscan apparatus (Molecular Imaging, Agilent) in the Acoustic AC Mode™. The images (topography and phase) where imported by ImageJ using a plugin available at http://rsb.info.nih.gov/ij/plugins/afm.html to analyze the features of the film surface quantitatively.

*Ellipsometry* was carried out in the Laboratoire de Physico-Chimie des Polymères et des Milieux Dispersés (ESPCI, Paris – France) with a SE400 Sentech Instrument apparatus operating at a wavelength $\lambda$ = 632.8 nm. The ellipsometric angles $\Psi$ and $\Delta$ were measured and fitted using the Fresnel equations of light reflection by an infinite silicon substrate (refractive indexes $n_S$ = 3.874 and $k_S$=0.0016) covered by 20 Å of $SiO_2$ oxide ($n_O$=1.46) and a layer of adjustable refractive index and thickness values. The measurement was repeated at 5 different locations on each wafer in order to get standard deviation values.

*Neutron Reflectivity (NR)* was performed in the Laboratoire Léon Brillouin (LLB, CEA Saclay – France) on the time-of-flight reflectometer EROS. The reflectivity curve was recorded at a constant angle 0.93° for a neutron wavelength $\lambda$ varying between 2.7 and 26 Å, thus covering a scattering vector range $q = 4\pi/\lambda \sin\theta$ between 0.078 and 0.75 nm$^{-1}$. The reflectivity curves were fitted with the SimulReflec software accessible at http://www-llb.cea.fr calculating a model reflectivity curve by iterations of the thicknesses $t_i$ and the coherent neutron scattering length densities (SLD) $Nb_i$. The initial configuration consisted in an infinite silicon medium ($Nb_{Si}$=2.14×10$^{-6}$ Å$^{-2}$) covered with a thin (1.3 nm) layer of oxide ($Nb_{SiO2}$=3.32×10$^{-6}$ Å$^{-2}$) and a succession of PS-PBMA/PBMA-PS bilayers ($Nb_{PS}$=1.41×10$^{-6}$ Å$^{-2}$ and $Nb_{PBMA}$=6.32×10$^{-7}$ Å$^{-2}$). An integer number of bilayers (*e.g.* 3, 4 or 5) were chosen according to the total film thickness measured independently by ellipsometry.

*Contact angle.* The free surface energy of the thin films was calculated by contact angle measurements, using the Owens-Wendt method with droplets of two liquids of different polarisabilities (water and ethylene glycol).[47]

## 3. Results and discussion

### 3.1 Synthesis of symmetric PBMA-b-PS diblock copolymer by ATRP

In this work, the requirements for the diblock copolymer synthesis were driven by the expected lamellar self-assembly of the copolymer within the film. At first, a similar degree of polymerization of each block had to be obtained ($N_{PS}$≈$N_{PBMA}$). Then the condition to reach the ordering transition writes $\chi N$ >10.5 for a symmetrical diblock copolymer, $\chi$ being the Flory-Huggins parameter describing the interaction of the two types of segments at the annealing temperature.[48] We estimate $\chi_{140°C}$=0.045 from a small angle neutron scattering study of binary blend mixtures of 10%-d(euterated)-PS and PBMA homopolymers[49] that leads to $\chi / V_0$=2.5×10$^{-4}$ Å$^{-3}$ (taking an average molecular volume $V_0$=($V_{PS}V_{PBMA}$)$^{1/2}$≈180Å$^3$). We know also $\chi_{160°C}$ = 0.037 from a value published for d-PS and PMMA,[50] a system close to ours. We deduce that the minimum degree of polymerization $N$ of each block should be around 300.

We first investigated the synthesis of the poly(n-butyl methacrylate) (PBMA) homopolymer by ATRP, which was subsequently used as the macroinitiator for styrene polymerization. The experimental conditions for synthesis of both PBMA macroinitiators and PBMA-*b*-PS diblock copolymers are reported in Table 1. The ATRP mechanism is based on the activation-deactivation equilibrium between dormant and active species which is established through a redox reaction using the copper-based catalyst. As depicted in equation 1, the ATRP kinetics of a specific monomer is mainly governed by the $k_P.K$ ratio, with $k_P$ the propagation rate constant and $K$ the equilibrium constant.

**Table 1.** Experimental conditions for the synthesis of PBMA macroinitiator and PBMA-*b*-PS diblock copolymers by ATRP. [a]

| Expt | Monomer | Initiator | Nature of catalyst | [M]:[Initiator]:[CuX]:[CuX$_2$]:[PMDETA] | Polym. time (h) |
|---|---|---|---|---|---|
| 1 | BMA | EBrIB | CuBr/CuBr$_2$ | 400:1:0.8:0.2:1 | 1.5 |
| 2 | BMA | EBrIB | CuCl/CuCl$_2$ | 400:1:0.8:0.21 | 2 |
| 3 | S | PBMA-Br (from expt 1) | CuCl | 665:1:1:1 | 15 |
| 4 | S | PBMA-Cl (from expt 2) | CuCl | 690:1:1:1 | 16 |

[a] Experimental conditions: T = 100°C; [BMA] = 6.2 mol.L$^{-1}$; [S] = 8.7 mol.L$^{-1}$

From the values of $k_p.K$ previously published[51] ($k_p.K$ = 1.1 × 10$^{-3}$ M$^{-1}$.s$^{-1}$ for MMA and 3.6 × 10$^{-5}$ M$^{-1}$.s$^{-1}$ for styrene polymerization at 90°C using EBrIB as initiator and 4,4'-(di-5-nonyl)-2,2'-bipyridine as copper ligand), the more reactive methacrylic monomer was logically polymerized as the first block in order to enhance the initiation efficiency of the macroinitiator for the second block, hence limiting the amount of residual first block.

$$\ln\frac{[M]_0}{[M]} = k_p.K.\frac{[Cu^{II}X_2][RX]}{[Cu^{I}X]}.t \qquad \text{Eq. (1)}$$

The ATRP of BMA was carried out at 100°C using 20 mol% of CuX$_2$ and 80 mol% of CuX as copper catalyst. The control over the BMA polymerization was observed whatever the nature of the copper halide (Br or Cl), as shown by low molar-mass dispersity indices ($M_w/M_n$ <1.2) and the correct matching between theoretical and experimental molar masses (Table 2).

**Table 2.** Analysis results of the PBMA homopolymer synthesis.

| Expt | Conversion (%) | $M_n$ theoretical [a] g.mol$^{-1}$ | $M_n$ SEC g.mol$^{-1}$ | $M_w/M_n$ | $f$ [b] |
|---|---|---|---|---|---|
| 1 | 87 | 49 680 | 57 520 | 1.19 | 0.86 |
| 2 | 84 | 47 975 | 60 860 | 1.18 | 0.79 |

[a] $M_n^{theoretical}=M_{EBrIB}+[BMA]_0/[EBrIB]_0\times conversion \times M_{BMA}$
[b] Initiator efficiency, $f = M_n$ theoretical / $M_n$ SEC.

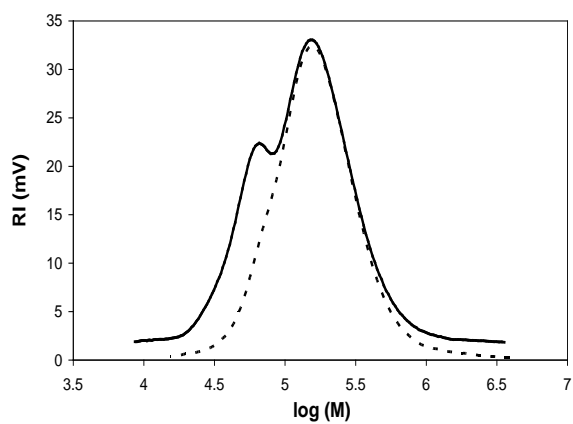
(a)

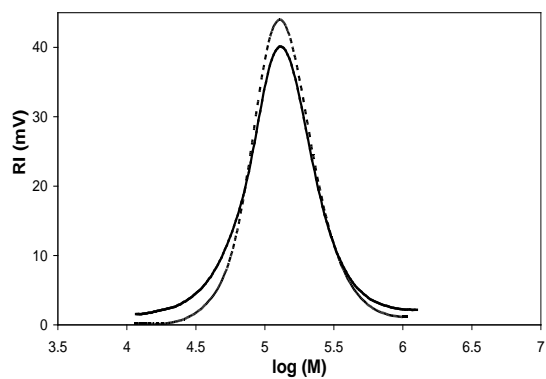
(b)

**Figure 1.** SEC of the PBMA$_{405}$-*b*-PS$_{460}$ (**a**, expt 3) and the PBMA$_{425}$-*b*-PS$_{490}$ (**b**, expt 4) diblock copolymers before (bold line) and after (dashed lines) hexane extraction.

Nevertheless, the nature of the copper halide catalyst affected the proportion of PBMA dead chains as highlighted by the SEC traces of the PBMA-*b*-PS diblock copolymers depicted in Figure 1. Indeed, the presence of residual PBMA homopolymer polluting the PBMA-*b*-PS diblock copolymer was noticeable when using the bromide-terminated PBMA macroinitiator (experiment 3) whereas pure diblock copolymer was recovered from the PBMA macroinitiator synthesized in the presence of the copper chloride catalyst (experiment 4). The higher proportion of PBMA dead chains is in accordance with the higher activation rate constant ($k_a$) of copper bromide-based catalyst in comparison with $k_a$ of the copper chloride-based catalyst.[52] The clear shift of the SEC trace of the diblock copolymer (Figure 2) together with the narrow molar mass distribution of the copolymer (Table 3) highlighted the control of the PBMA-*b*-PS diblock copolymer synthesis up to high molar masses (≈ 100 000 g.mol$^{-1}$). In conclusion, we successfully synthesized well-defined copolymers PBMA$_{405}$-*b*-PS$_{460}$ (exp. 3) and PBMA$_{425}$-*b*-PS$_{490}$ (expt 4) which fulfilled the initial requirements, *i.e* the synthesis of a symmetrical diblock copolymer with an overall degree of polymerization up to 915.

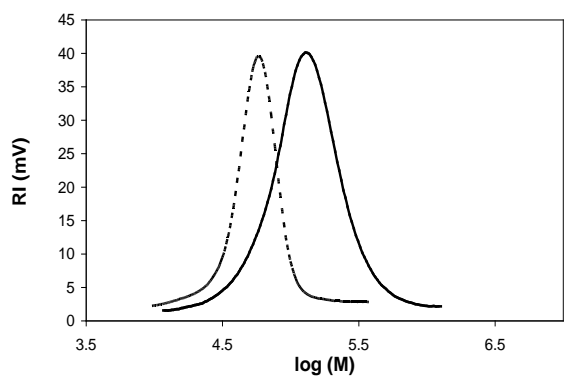

**Figure 2**. SEC of the PBMA$_{425}$ macroinitiator (dashed lines, expt 2) and the PBMA$_{425}$-*b*-PS$_{490}$ diblock copolymer (bold line, expt 4).

**Table 3.** Analysis results of the PBMA-*b*-PS diblock copolymer synthesis.

| Expt | Conversion (%) | $M_{n\ theoretical}$ [a] g.mol$^{-1}$ | Mn SEC [b] g.mol$^{-1}$ | $M_n$ SEC [c] g.mol$^{-1}$ | $M_w/M_n$ [c] | f [d] | Mn NMR [e] g.mol-1 |
|---|---|---|---|---|---|---|---|
| 3 | 53 | 94 175 | 85 400 | 105 400 | 1.38 | 0.9 | 102 740 |
| 4 | 77 | 116 115 | 102 400 | 111 900 | 1.44 | 1.0 | 114 250 |

[a] $M_n^{theoretical}=M_n(PBMA)+[S]_0/[PBMA]_0\times conversion\times M_S$

[b] $M_n$ of the crude diblock copolymer.

[c] Macromolecular characteristics of the di-block copolymer recovered after extraction of the residual PBMA homopolymer with hexane.

[d] $f$ is the initiator efficiency: $f = M_n^{theoretical}/M_n^{SEC}$ using $M_n^{SEC}$ of the diblock copolymer free of any residual PBMA macroinitiator.

[e] $M_n^{NMR}=M_{EBrIB}+DP_{BMA}\times M_{BMA}+DP_S\times M_S$, with DP the average degree of polymerization. D$_{PBMA}$ was calculated using $M_n^{SEC}$ of the macroinitiator and DP$_S$ was calculated on the basis of the proton NMR integrations of both monomer units.

### 3.2 Synthesis of γ-Fe$_2$O$_3$@PS nanoparticles

This study necessitated to synthesize magnetic nanoparticles and disperse them as inorganic fillers in the targeted nanocomposite films. Those MNPs could be brought either as a dry powder or as a ferrofluid, *i.e* a stable colloidal dispersion. Our aim was to obtain nanocomposite films well ordered as lamellae with the inorganic fillers embedded in some of the layers only. This purpose could not be reached by starting from a dry powder because irreversibly aggregated nanoparticles would not disperse properly into polymer lamellae. To confine MNPs within the layers of the diblock copolymer and keep a lamellar order, it was necessary to coat those nanoparticles with a thin polymer layer of the same nature as one block of the copolymer. This strategy to obtain nanocomposite materials was once called *ex-situ synthesis* of the MNPs *and co-assembly* with the BCPs.

Among the different available magnetic nanoparticles, we focused our attention on those made of γ-Fe$_2$O$_3$ (maghemite), which is a pure iron$^{+III}$ oxide that belongs to the spinel crystallographic structure. The TEM pictures showed "rock-like" nanoparticles, which can be approximated as spheres having diameters between 5 and 12 nm (see Figure 3). Their size distribution was rather broad as shown by the Log-normal fit with median value $d_0$ = 6.8 nm and standard width $\sigma$ = 0.39, as determined by VSM (ESI-Figure 1). In this work, we also examined the advantage of using MNPs with a fairly narrower size distribution. A size sorting procedure by successive phase separations enabled to decrease the width down to $\sigma$ = 0.25 for a median $d_0$ = 6.2 nm.[39] The TEM micrographs show the visible difference between unsorted "polydisperse" and size-sorted "monodisperse" γ-Fe$_2$O$_3$ nanoparticles (Figure 3).

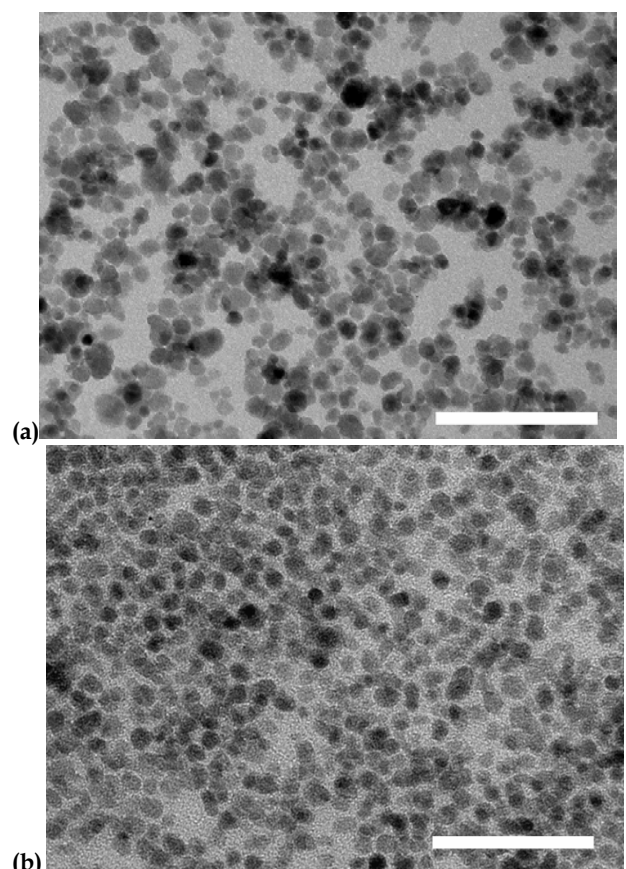

**Figure 3.** TEM micrographs of polydisperse **(a)** and monodisperse **(b)** γ-Fe$_2$O$_3$ nanoparticles. The scale bar length is 50 nm in both cases.

**Table 4.** Experimental conditions to synthesize γ-$Fe_2O_3$@PS MNPs by SI-ATRP (at 100°C).

| Expt. | γ-$Fe_2O_3$ $d_0$ (nm) / $\sigma^a$ | [Styrene] mol.L$^{-1}$ | [St]/[ γ-$Fe_2O_3$@BrMPA]/ [CuBr]/[PMDETA] | Polym. time (min) |
|---|---|---|---|---|
| γ-$Fe_2O_3$@PS1 | 6.8 / 0.39 | 7.84 | 20:1:1:1 | 30 |
| γ-$Fe_2O_3$@PS2 | 6.2 / 0.25 | 7.89 | 20:1.5:1:1 | 30 |
| γ-$Fe_2O_3$@PS3 | 7.0 / 0.23 | 7.30 | 10:1.5:1:1 | 45 |

$^a$ Two parameters of a Log-normal distribution for the magnetic core's diameters obtained by fitting the VSM curves with Langevin's law.

**Table 5.** Analysis results of the γ-$Fe_2O_3$@PS nanoparticles synthesis.

| Expt. | dh (nm)$^a$ | Mn PS SEC$^b$ (g.mol$^{-1}$) | Mw/Mn $^b$ | % w/w γ-$Fe_2O_3$$^c$ | % w/w PS$^d$ |
|---|---|---|---|---|---|
| γ-$Fe_2O_3$@PS1 | 144 | 45 820 | 2.45 | 27 | 73 |
| γ-$Fe_2O_3$@PS2$^e$ | 210 / 60 | 23 840 | 1.98 | 1.3 | 98.7 |
| γ-$Fe_2O_3$@PS2$^f$ | 280 / 50 | 19 330 | 1.77 | 7.4 | 92.6 |
| γ-$Fe_2O_3$@PS3 | 130 | 15 560 | 1.41 | 79 | 21 |

$^a$ Hydrodynamic diameter of filtrated solutions measured by DLS (two values indicate a bimodal distribution).
$^b$ Macromolecular data obtained by SEC after separation from iron oxide (see 2.3 and 2.4).
$^c$ Obtained after titration of iron by absorption spectrometry.
$^d$ Deduced by difference % w/w PS = 100% -% w/w γ-$Fe_2O_3$.
$^e$ CuBr removed by filtration after polymerization.
$^f$ CuBr removed by water rinsing.

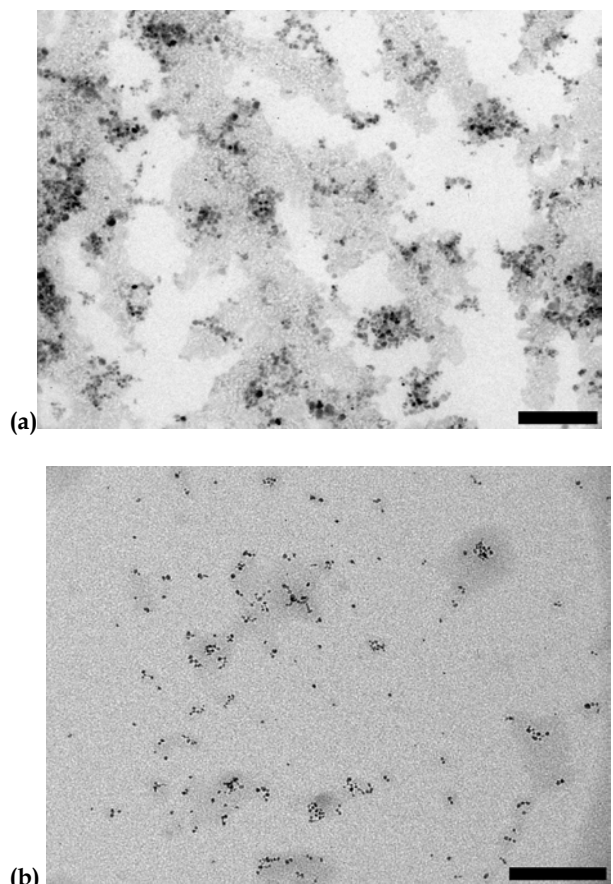

**Figure 4.** TEM micrographs of "polydisperse" γ-$Fe_2O_3$@PS1 **(a)** and "monodisperse" γ-$Fe_2O_3$@PS2 **(b)** nanoparticles. The scale bar length is 200 nm in both cases. The grey areas around the black inorganic cores are obtained by coloration of the PS aromatic rings by $RuO_4$.

We used those two types of nanoparticles to investigate a possible effect of the particle size distribution on the control of polymerization during the "grafting from" synthesis to get γ-$Fe_2O_3$@PS nanoparticles. The experimental conditions for three syntheses of PS-grafted nanoparticles are reported in Table 4. Because of washing steps after the ligand-exchange, the presence of free initiator in excess is unlikely. Thus the actual molar ratio of initiator *vs.* catalyst might be smaller than the theoretical values of 1 or 1.5.

After synthesis, we obtained large aggregates of core-shell nanoparticles ($d_h$ > 150 nm) that required filtration to obtain smaller aggregates (Table 5). Nevertheless, the TEM micrographs of those particles showed that the broadly distributed maghemite nanoparticles led to large aggregates, while the "monodisperse" nanoparticles led to much fewer small aggregates in coexistence with isolated nanoparticles coated by PS corona (Figure 4). This aggregation effect observed during the SI-ATRP polymerization of styrene may be ascribed to termination reactions between growing chains initiated from the functionalized nanoparticles surface. According to Matyjaszewski, such termination reactions very probable in bulk-polymerization could be eliminated by performing SI-polymerization in a dilute dispersed phase instead[51] or under high pressure around 6 kbar.[53]

Furthermore, monodisperse nanoparticles γ-$Fe_2O_3$@PS3 gave a better-defined SEC trace after de-grafting of the chains than the polydisperse ones γ-$Fe_2O_3$@PS1 (Figure 5). This indicates that a narrow distribution of diameters of the magnetic cores is somehow related to the quality of CRP and the final state of aggregation. In the literature, several studies about SI-ATRP at the surface of iron oxide MNPs clearly evidenced the living character of polymerization but without leading systematically to a low degree of aggregation for the core-shell products. A. Kaiser *et al.* used polydisperse $Fe_3O_4$ (magnetite) MNPs with a size distribution analogous to our sample 1 and also coated with BrMPA for the SI-ATRP of styrene in toluene at three monomer concentrations.[36] In spite of the heterogeneous phase (the samples exhibiting a culot and a supernatant with $d_h \approx 200$ nm), molar-mass dispersity indices were always below 1.3 with $M_n$ up to 90000 g.mol$^{-1}$. The authors showed that the aggregation of MNPs clearly reduces the level of available initiator, providing PS chains of higher molar masses.

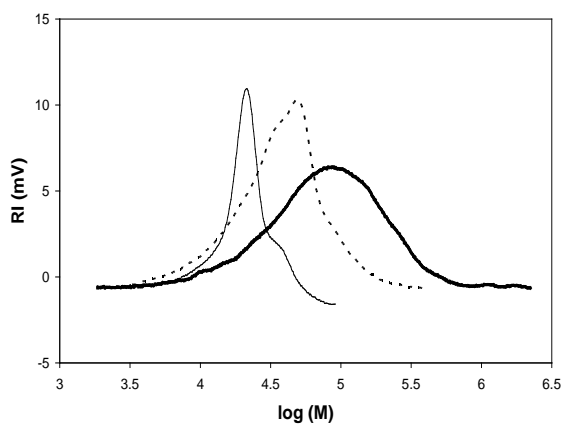

**Figure 5.** SEC curves of the polymer chains after degradation of the inorganic cores of γ-Fe$_2$O$_3$@PS1 (bold line), γ-Fe$_2$O$_3$@PS2 (dashed lines) and γ-Fe$_2$O$_3$@PS3 (thin line).

In the present work, we observe also that a higher experimental $M_n$ is concomitant with a broader molar mass distribution $M_w/M_n$ (Table 5). Therefore a lower concentration of grafted initiator is involved to create chains, producing a lower amount of control agent (*i.e.* Cu(II)Br$_2$) by the so-called "persistent radical effect".[54] Another reported study strengthens the idea that a lower size-dispersity of the magnetic cores improves the control of radical polymerization: S. M. Gravano *et al.* used indeed maghemite γ-Fe$_2$O$_3$ MNPs made by iron carbonyl decomposition,[34] for which we estimate that the width parameter $\sigma$ of a Log-normal distribution is below 0.1, to compare with 0.2 at best for MNPs synthesized by aqueous route. Having grafted a 2-bromopropionate ester on them, the authors studied the SI-ATRP of styrene in the bulk on those highly monodisperse magnetic cores. The perfect 1st-order and living character of the polymerization being evidenced by the authors, the PS chains were short and controlled ($M_n$<28000 g.mol$^{-1}$ for 50% conversion) with a particularly narrow molar mass distribution ($M_w/M_n$<1.1). However, this article exhibits a TEM image of large (hundreds of nm) aggregates of MNPs embedded in a PS matrix looking like our Figure 4a.

In conclusion, we synthesized γ-Fe$_2$O$_3$@PS core-shell nanoparticles by surface-initiated polymerization (also called "grafting from") using ATRP. Although aggregates persisted, we slightly improved the control of the radical polymerization (compared to our first attempt on Table 5) and decreased the proportion of aggregates by using iron oxide nanoparticles more homogeneous in sizes for the SI-ATRP synthesis. Even though more efforts would be necessary to explain quantitatively a surface curvature effect when nano-sized particles are used for SI-ATRP

(*e.g.* by measuring the grafting density of BrMPA initiator on several batches of MNPs of different distributions of diameters), the evidenced relationship between the characteristics of inorganic synthesis (distribution of diameters) and of polymer synthesis (molar mass dispersity indexes) adds valuable information about that domain.

*3.3. Formation and characterization of pure copolymer films*
*Lamellar order of pure copolymer films.*

At first, we studied pure copolymer films without nanoparticles in order to check their lamellar order. According to the literature, films made of symmetrical copolymers annealed in a range of temperatures above the $T_g$ of their blocks and between the upper (UCOT) and lower (LCOT) critical ordering transitions[2] self-assemble into regular lamellae. Such an organization leads to characteristic defects at the film surface because the top layer is rarely complete. Depending on the initial thickness of deposited copolymer (never strictly equal to an integer number of the lamellar period), these defects appear either as "islands" or "holes".[11, 12]

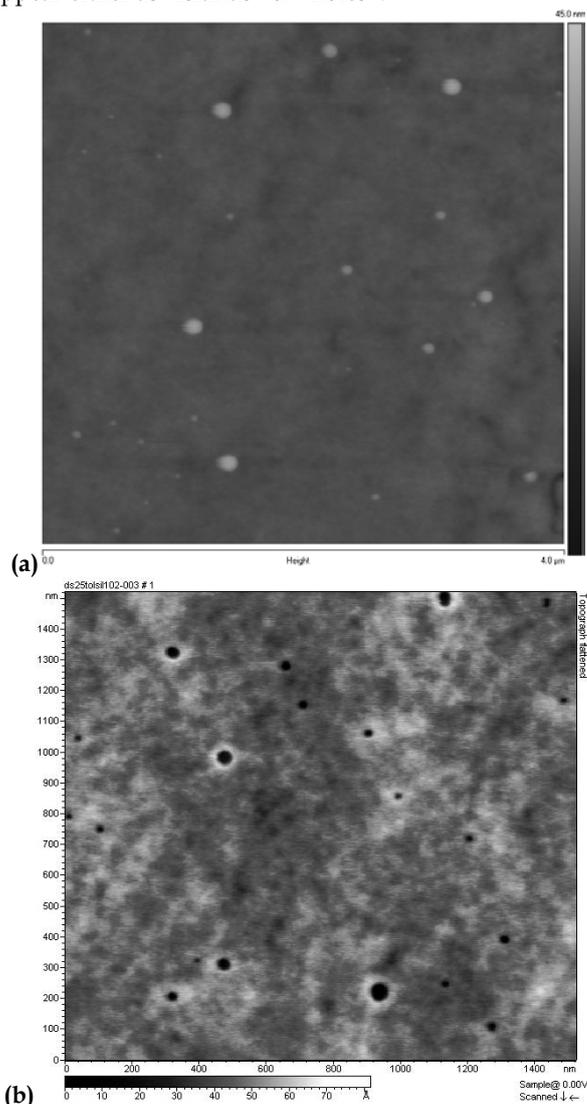

**Figure 6.** AFM height images of pure PBMA425-*b*-PS490 films leading either to "islands" **(a)** or "holes" **(b)** at their top surface depending on thickness. Statistical analyses of the heights, depths and diameters of those surface defects are provided as ESI-Figure 2.

Both types of characteristic defects were detected by AFM with PBMA$_{425}$-*b*-PS$_{490}$ films by varying the deposited volume and substrate area (Figure 6). In the case of the islands, the statistical analysis of this AFM topographical picture provided us an average height $H_0$=16.37±0.73 nm that we can identify with half the value of the lamellar spacing[11, 12, 15, 24, 55] to get $L_{AFM}$=32.7±1.5 nm. Ellipsometry data gave us the global thickness of the films with an indication about their roughness (standard deviation), *e.g.* 122.5±1.1 nm for the undoped film (nb. 2 on Table 6). The average optical refractive index *n* =1.540±0.004 for the pure copolymer is reasonable for a thin organic layer on silicon.

Neutron reflectivity (NR) enabled us to study the system at a much more mesoscopic scale. A flat un-structured film such as ours right after spin-coating gives a reflectivity profile that decreases globally as $q^{-4}$ while being modulated by a regular oscillation (Kiessig fringes) with a constant period related to the film thickness. After annealing, we observed an overshoot (*i.e.* an amplification of the first maximum compared to the following ones) which is a direct evidence of the lamellar order (Figure 7(a)).[57] Considering this highest peak at a wave vector $q_{max}$≈0.166 nm$^{-1}$ as due to quasi-Bragg reflection on the lamellae, we estimated the average lamellar period by $L_{ave}$=2π/$q_{max}$≈37.9 nm for the undoped film, which is 16% larger than from AFM. For such nanostructured films, the super-oscillations of the NR curve replacing the regular Kissieg fringes before annealing enabled to build a scattering length density (SLD) profile with a total thickness of the film of $L_{tot}$=$L_1$+$L_2$+$L_3$=118 nm in accordance with ellipsometry and with quite a regular period (Figure 7(b)).

**Table 6.** Analysis results by ellipsometry and NR of the thin films of PBMA$_{425}$-*b*-PS$_{490}$ at a volume fraction *Φ* in MNPs after annealing at 150°C.

| Film nb. | Ann. time (h) | *Φ* (v/v) γ-Fe$_2$O$_3$ | Thickness[a] (nm) | RI[a] | $L_1$[b] (nm) | $L_2$[b] (nm) | $L_3$[b] (nm) | $L_4$[b] (nm) | $L_5$[b] (nm) | $L_{tot}$[c] (nm) | $L_{ave}$[d] (nm) | $N_{layer}$= $L_{tot}$/$L_{ave}$ |
|---|---|---|---|---|---|---|---|---|---|---|---|---|
| 2 | 48 | 0 | 122.5 ± 1.1 | 1.540 ±0.004 | 30.0 | 42.2 | 45.4 | - | - | 117.6 | **37.9** | 3.1 |
| 5 | 48 | 0.025% | 114.6 ± 0.8 | 1.562 ±0.005 | 31.2 | 45.3 | 41.5 | - | - | 117.9 | **39.3** | 3.0 |
| 8 | 72 | 0.05% | 108.1 ± 1.4 | 1.642 ±0.013 | 32.2 | 48.0 | 31.5 | - | - | 111.7 | **42.2** | 2.6 |
| 9 | 72 | 0.10% | 109.4 ± 3.5 | 1.607 ±0.023 | 32.8 | 48.0 | 33.3 | - | - | 114.0 | **41.1** | 2.8 |
| 10 | 72 | 0.15% | 106.6 ± 0.4 | 1.660 ±0.017 | 32.8 | 48.7 | 30.0 | - | - | 111.5 | **43.8** | 2.5 |
| 11 | 72 | 0.25% | 116.6 ± 5.2 | 1.606 ±0.030 | - | - | - | - | - | - | - | - |
| 12 | 72 | 0.025% | 50.2 ± 0.7 | 1.533 ±0.007 | - | - | - | - | - | - | - | - |
| 13 | 72 | 0.025% | 184.0 ± 4.9 | 1.536 ±0.004 | 34.0 | 46.1 | 49.0 | 54.5 | - | 183.6 | 41.9 | 4.4 |
| 14 | 72 | 0.025% | 248.4 ± 17.4 | 1.727 ±0.097 | 45.4 | 46.3 | 46.6 | 46.2 | 44.4 | 228.9 | 41.1 | 5.7 |

[a] Thickness, roughness (standard deviation) and refractive index measured by ellipsometry.
[b] $L$i is the thickness of the i$^{th}$ bilayer PS-*b*-PBMA/PBMA-*b*-PS as measured from the SLD profile fitting the neutron reflectivity curve.
[c] Total film thickness deduced from the fitting SLD profile $L_{tot}$ =Σ$L_i$ .
[d] Average bilayer thickness deduced from the quasi-Bragg peak $L_{ave}$=2π/$q_{max}$ of the NR curve. The five values in bold cases can be fitted by a linear regression $L_{ave}$=$L_0$(1 + *pΦ*) with $L_0$=38.6 nm and *p*=89 as described by B. Hamdoun *et al*.[16, 56]

Considering the thickness of the film and the average lamellar period, we can calculate the average number of bilayers by $N_{layer}$=$L_{tot}$/$L_{ave}$. This number being close but slightly above an integer (here $N_{layer}$=3.1), the top surface of this film exhibit island-type defects on the AFM picture of Figure 6(b). In addition, the average lamellar period of this film is about 38 nm *i.e.* 31% larger that the value 29.5±0.5 nm measured in previous studies with films prepared with a PS-*b*-PBMA sample of molar mass $M_w$=82000 g.mol$^{-1}$ synthesized by anionic polymerization with $M_w$/$M_n$=1.05 that was used for several structural studies of incorporation of MNPs.[15-18] Taking into account the 46% increase of $M_n$ for our PBMA425-*b*-PS490 synthesis compared to this reference sample, one calculate a scaling law *L*ave~$M_n^{0.67}$. It may be fortuitous, but this value 31%/46%=0.67 is close to the theoretical exponent 0.643 for the scaling law of the lamellar period of symmetrical BCPs *vs.* molar mass[58] that was firstly studied experimentally by Hadziioannou *et al.* who measured a larger exponent (0.79) for PS-*b*-PI copolymers.[59] We gathered our measurement with all the published values of the lamellar period of (sometimes deuterated) PS-*b*-PBMA films of $M_n$ varying from 64000 to 650000 g.mol$^{-1}$ measured by several methods (Small Angle X-ray[2, 25] and Neutron[13] Scattering, X-ray[60] and Neutron Reflectivity[14, 17, 18, 23, 24], AFM[15, 24, 55] and optical interference microscopy[12]) on a master curve (ESI-Figure 3). This bibliographic study leads to a scaling law exponent around 0.69, very close to the exponent 0.643 predicted by Helfand due to compaction of the chains in the lateral direction into dense bilayers under surface tension and thus stretching in the direction normal to the lamellae.[58] For unilamelar vesicles with hydrophopic amorphous blocks like polybutadiene, which are a particular case of the lamellar state of BCPs, Eisenberg showed that the scaling exponent of bilayer thickness *vs.*

molar mass varies from 1/2 for the lower masses (*e.g.* <10000 g.mol$^{-1}$) to 2/3 for the larger ones.[61] With PBMA-*b*-PS around 100000 g.mol$^{-1}$, we are certainly in the latter regime and thus the chains are more extended that the conformation of random coils hypothesized by Hamdoun in his theoretical model of nanocomposite BCP/MNP thin films.[62]

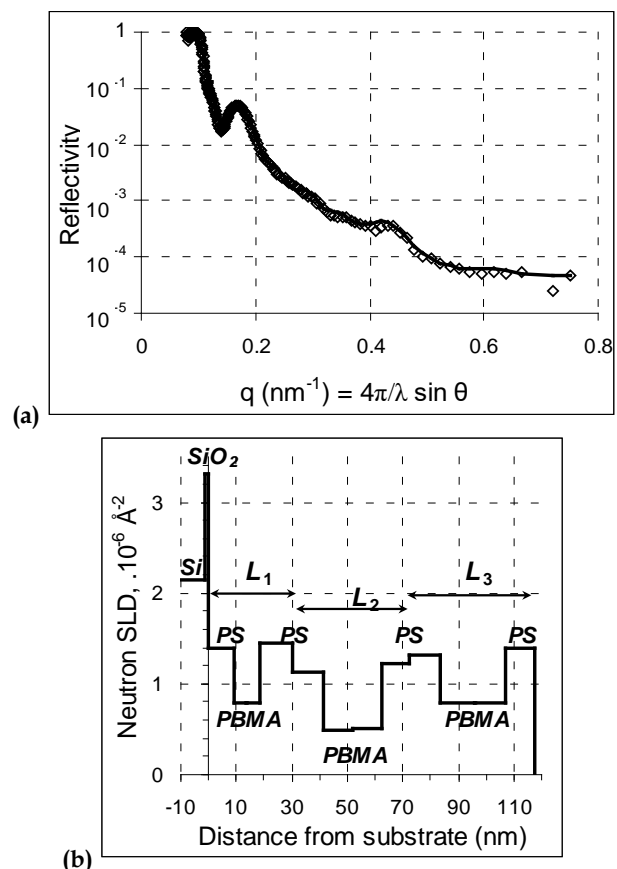

**Figure 7.** **(a)** Neutron Reflectivity spectrum of a PBMA$_{425}$-*b*-PS$_{490}$ copolymer film and **(b)** calculated SLD profile deduced from the fitting curve.

*Interpenetration and sequence order of the blocks.* NR provides other information in addition to the lamellar period. The calculated neutron SLD profile on Figure 7(b) enables superposing a calculated reflectivity curve (solid line) to the experimental points of the pure film (diamonds on Figure 7(a). The lower contrast between the calculated $Nb_{PBMA}$ (increase) and $Nb_{PS}$ (decrease) compared to their initial values before curve fitting (pure homopolymers) shows that the copolymer blocks may not be totally phase-separated. By looking at the SLD profile, one notices also that the first bilayer thickness $L_1$ (the closest to silicon) is significantly lower than the two following ones, an effect ascribed to interpenetration of the PBMA blocks. More importantly, the sequence order of the alternating PS and PBMA blocks in the lamellae is different from the previously reported systems.[17, 18, 23] Contrary to these studies showing the PBMA blocks always at the interfaces due to a lower surface energy, all our SLD profiles correspond to the PS blocks in contact with both the air and the substrate. The opposite hypothesis for the alternating layers starting from the silicon chip with PBMA blocks led systematically to a lower fitting quality of the simulations to the experimental NR curves. We ascribe this different sequence order to the modification of chemical end-groups of the copolymers used here. In the previous studies indeed, the diblock PS-*b*-PBMA copolymers were always synthesized by anionic polymerization. In the present work, the PBMA-*b*-PS copolymer was synthesized by ATRP, implying a halide initiator / catalyst system and therefore different chemical end-groups. Contact angle measurements were carried out to test this explanation by calculating the interfacial energy of those films (top solid surface/air). Films of either pure PBMA (expt 2 in Tables 1 and 2) or PS (also synthesized by ATRP) homopolymers and of the PBMA$_{425}$-*b*-PS$_{490}$ copolymer were prepared in the same conditions as all films studied by NR (in particular the same amount of deposited polymer onto the substrate). While the values of surface energy with air were similar for the PS homopolymer (66.01 mJ.m$^{-2}$) and for the diblock copolymer (63.53 mJ.m$^{-2}$), the surface energy was much lower for the PBMA homopolymer film (7.46 mJ.m$^{-2}$). In the case of PS-*b*-PMMA synthesized by anionic polymerization, only a slight difference of surface energies (less than 1%) was reported between the PS and the PMMA blocks, which enables a fine tuning of the anchoring of the lamellae between parallel and perpendicular orientation to the susbrate[9, 63] For PBMA-*b*-PS prepared by ATRP, there is on the contrary a strong contrast of surface energies between the two blocks. Those wetting measurements confirm the profiles deduced from neutron reflectivity with PS layers of the diblock copolymer in contact with air. This phenomenon is counter-intuitive when one takes into account only the surface energy with air which is much lower for PBMA. As shown indeed for blends of PS and PBMA homopolymers in the same range of molecular weights (around 10$^5$ g.mol$^{-1}$), the interface is enriched in PBMA compared to bulk and the top surface is always made of a thin layer (below 1 nm) of pure PBMA, as expected from its lowest surface tension with air.[64] In the case of a thin film of diblock PBMA-b-PS copolymer, the total interfacial energy contains not only the surface tension with air but also the interfacial tension with the substrate. Considering the ATRP route chosen to synthesize our copolymer (Table 1 and ESI-Scheme 1), we expect that brominated or chlorinated chemical groups are located at the end of the PS block (but not of the PBMA one that was polymerized first). Therefore a polarisability increase might be sufficient to lower the interfacial energy between the silicon substrate and the PS layers compared to the PBMA ones. Once the first block in contact with Si is determined by higher polar interactions, the nature of the top surface layer is constrained by the total thickness $L$tot of the film. This scenario might explain the inversion of the nature of the blocks in contact with the wafer and with air compared to previously published results.

*3.4. Formation and characterization of films doped with magnetic nanoparticles*

Similar analyses by AFM, NR and ellipsometry were carried out with PBMA$_{425}$-*b*-PS$_{490}$ thin films doped with MNPs from the "polydisperse" batch γ-Fe$_2$O$_3$@PS1. Five values of the iron oxide volume fraction were studied for increasing amounts of nanoparticles between $\Phi$=0.025% and $\Phi$=0.25% (films nb. 5, 8, 9, 10, 11 on Table 6). The lamellar order was preserved up to $\Phi$ = 0.15%, as evidenced by the super-oscillation (overshoot) of the NR curves typical of a lamellar period (Figure 8). A volume fraction of iron oxide in the film $\Phi$ = 0.25% was above the threshold value at which the lamellar order becomes distorted.

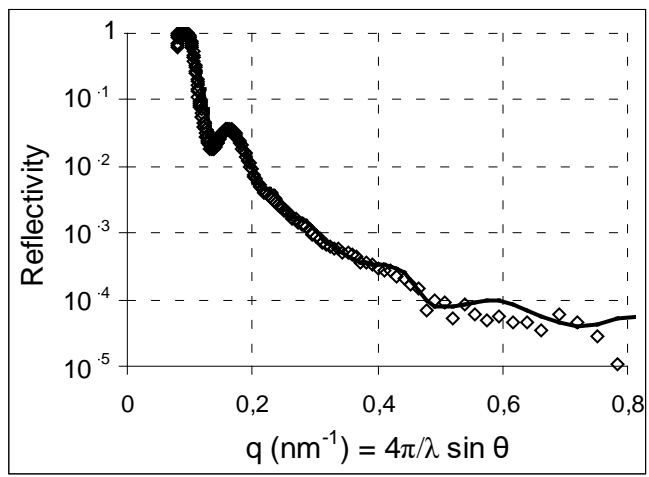

(a1)

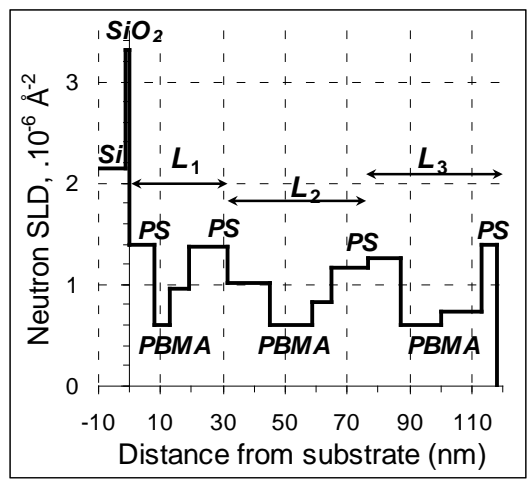

(b1)

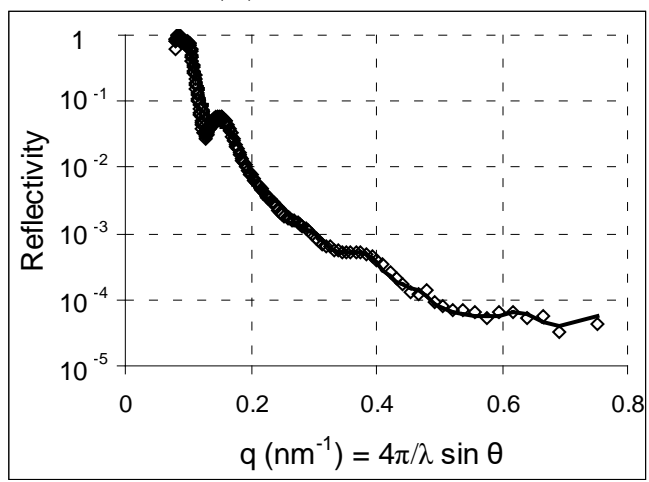

(a2)

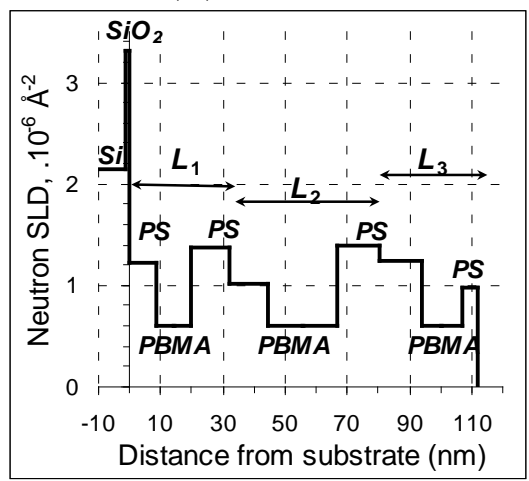

(b2)

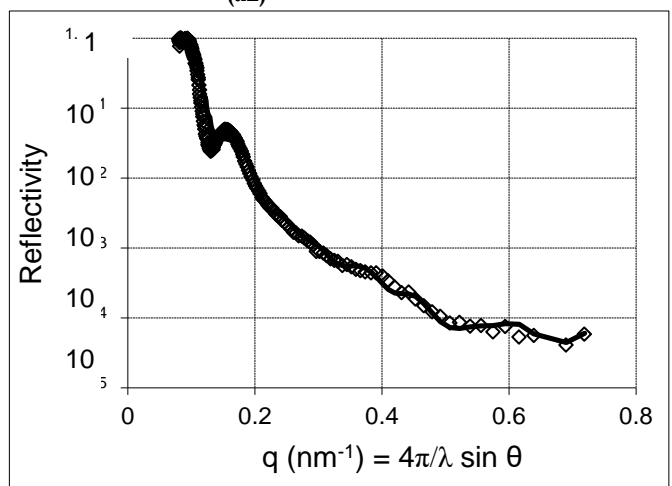

(a3)

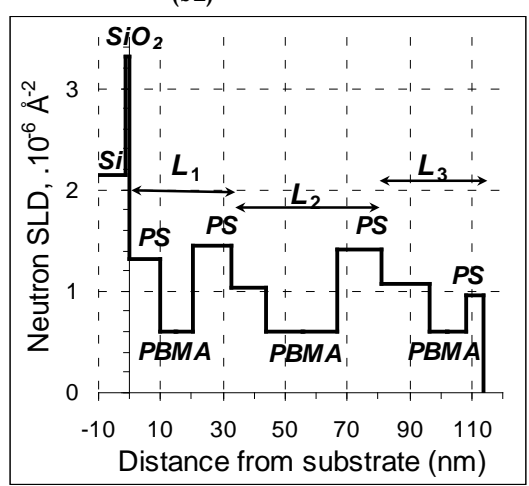

(b3)

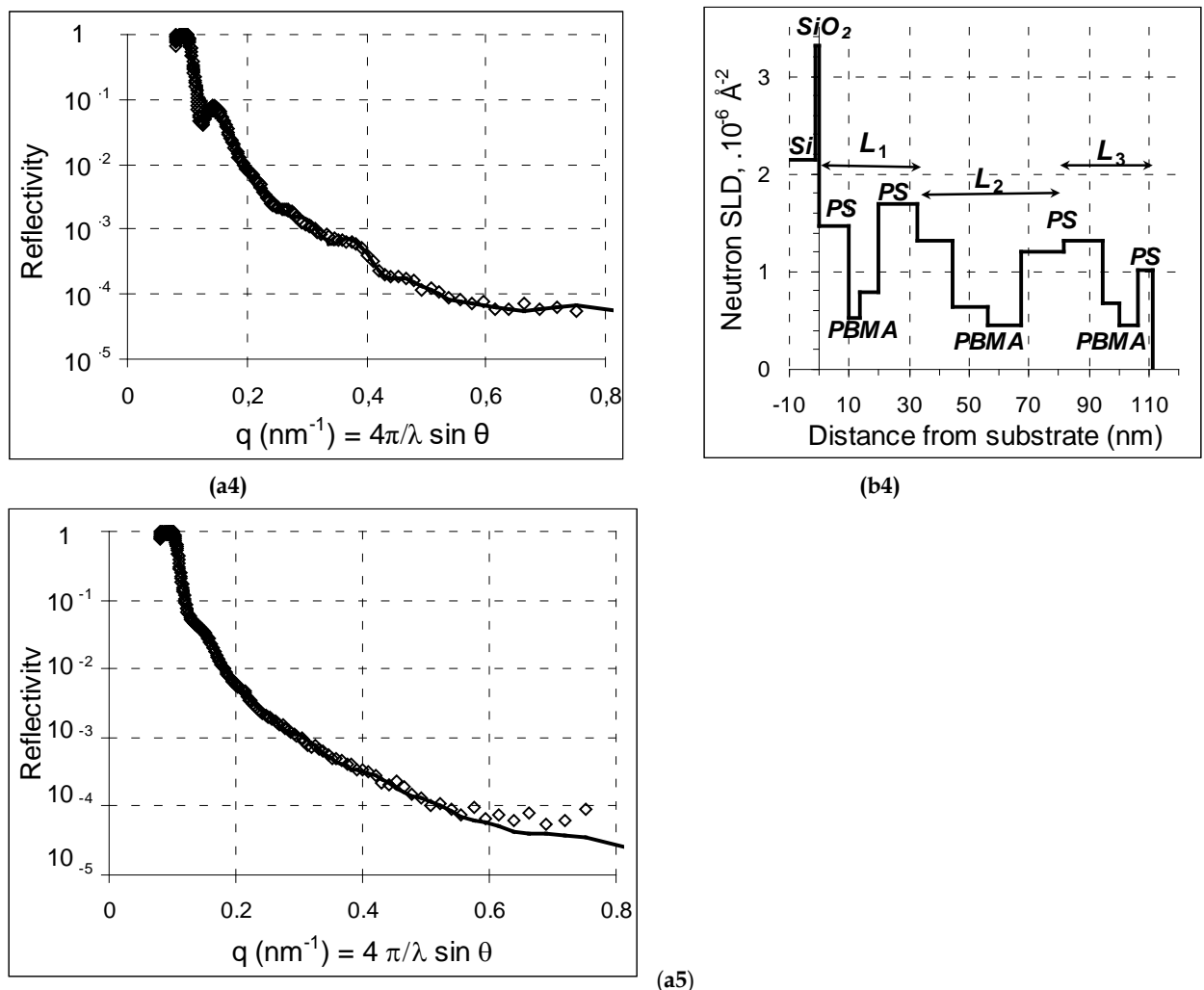

**Figure 8. (ai)** Neutron Reflectivity curves (raw data and fitting curves) and **(bi)** neutron SLD profiles deduced from the fit for PBMA$_{425}$-*b*-PS$_{490}$ films doped with γ-Fe$_2$O$_3$@PS1 nanoparticles prepared by spin-coating on silicon of 750µL of a toluene solution of block copolymer (BCP) and magnetic nanoparticles (MNP) at total concentration $C$ = 20 g/L with increasing volume fractions $\Phi$ (γ-Fe$_2$O$_3$/γ-Fe$_2$O$_3$+BCP) between 0 and 0.25% (for each value, the annealing time at 150°C is also given): **(a1, b1)** $\Phi$ = 0.025% (48h); **(a2, b2)** $\Phi$ = 0.05% (72h); **(a3, b3)** $\Phi$ = 0.10% (72h); **(a4, b4)** $\Phi$ = 0.15% (72h); **(a5)** $\Phi$ = 0.25% (72h).

A complementary study was made by keeping the volume fraction of γ-Fe$_2$O$_3$ at $\Phi$ = 0.025 % and varying the deposited concentration $C$ in order to check the influence of the total film thickness on nanostructuration (Table 6 films nb. 12, 13, 14 and corresponding NR curves on Figure 9).

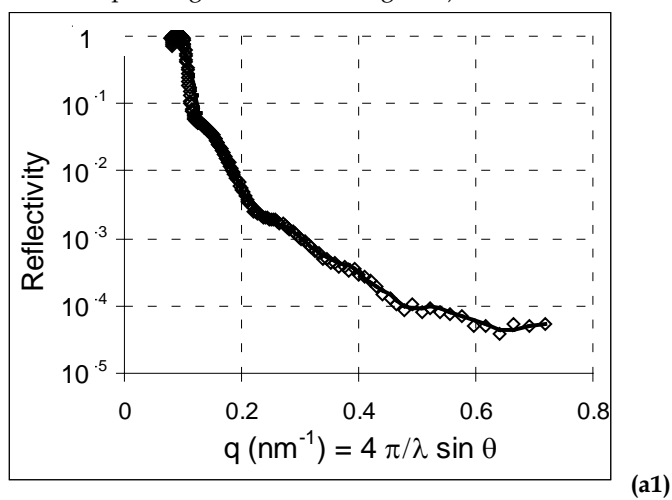

(a1)

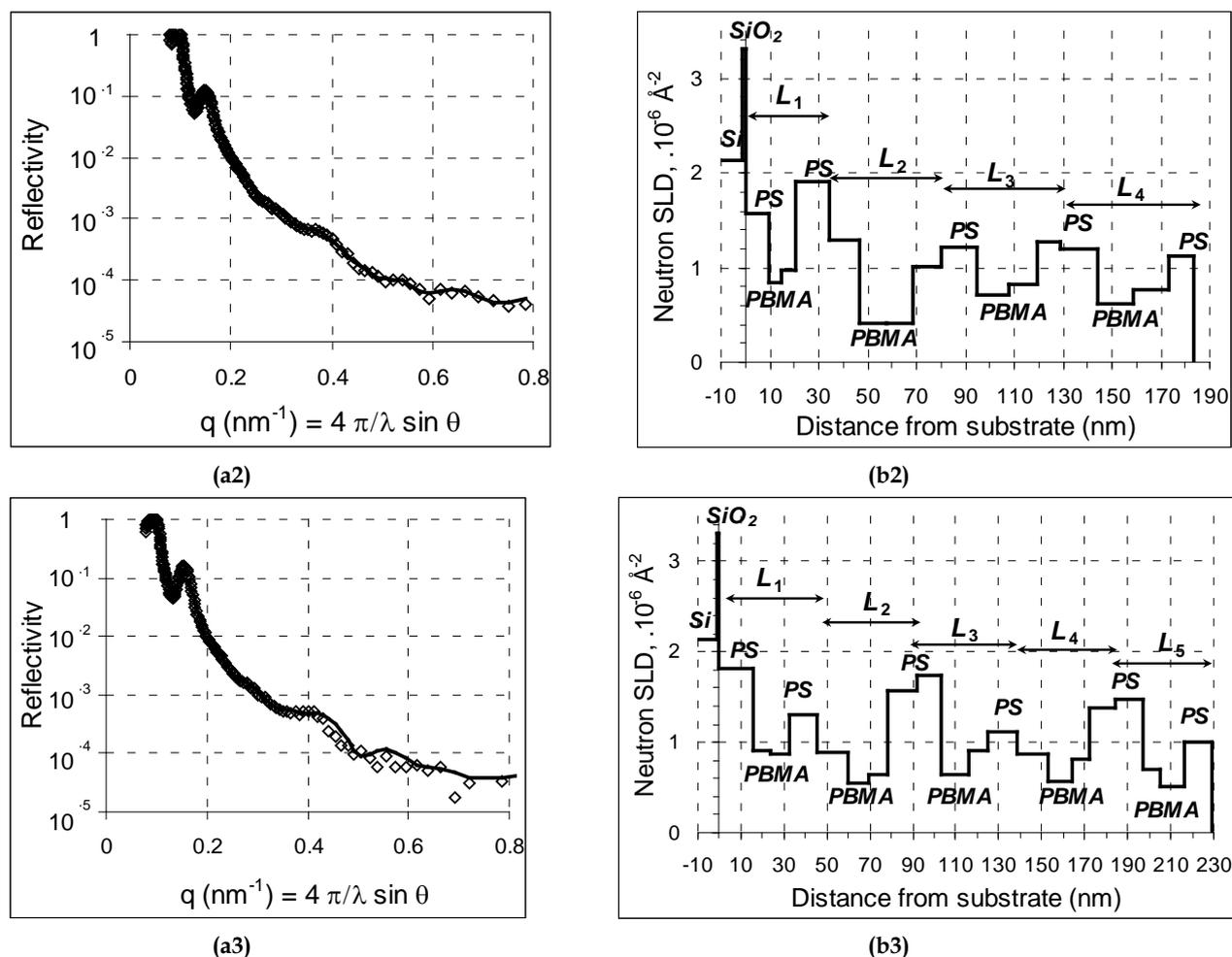

**Figure 9. (ai)** Neutron Reflectivity curves (raw data and fitting curves) and **(bi)** neutron SLD profiles deduced from the fit for $PBMA_{425}$-$b$-$PS_{490}$ films doped with $\gamma$-$Fe_2O_3$@PS1 nanoparticles at $\Phi$ = 0.025% for varying total concentration $C$ of the solution deposited on silicon (annealing time at 150°C is always 72h): **(a1)** $C$ = 10 g/L ; **(a2, b2)** $C$ = 30 g/L ; **(a3, b3)** $C$ = 40 g/L.

*Swelling of the lamellar period by the PS-functionalized MNPs.*

As long as the lamellar order was preserved, the incorporation of MNPs into the films had a visible effect on the quasi-Bragg peak that was shifted to slightly lower $q_{max}$ values. This corresponds to an almost monotonous increase of the average lamellar period in the doped case from $L_{ave}$=39.3 nm for the minimum volume fraction $\Phi$=0.025% to $L_{ave}$=43.8 nm for the maximum one $\Phi$=0.15%. Following the model by B. Hamdoun et al.,[16, 56] we compute a linear regression $L_{ave}$=$L_0(1+p\Phi)$. With this formalism, a value $p$=1/3 corresponds to the case of "small nanoparticles" (*i.e.* entropy driven) confined between adjacent PS layers, while "large nanoparticles" that swell the PS blocks homogeneously would yield $p$=1. In our study, we find $p$≈89, which is much larger than the experimental value $p$=0.27 reported for lamellar nanocomposites prepared from PS-$b$-PBMA and MNPs coated by PS chains both synthesized by anionic polymerizations.[16, 17] This discrepancy might originate from the much higher organic content of the $\gamma$-$Fe_2O_3$@PS nanoparticles in our syntheses (Table 5). With a fraction 73 % w/w of PS chains for our "polydisperse" sample, we calculate that the magnetic cores represent only 7.1% v/v of the MNPs. Thus we can recalculate the true volume fractions of the MNPs including their organic shells $\Phi_{cor}$≈$\Phi$/0.071 for all the doped films. Within this correction, the slope becomes $p_{cor}$=6.3 which is still above 1, maybe as an effect of the slight aggregated state of the MNPs. By looking carefully at the NR measurements of the doped films, we remark that the second step of the SLD profiles is almost always above the value of pure PS (except for the thickest film nb. 14 were the maximal value of $Nb$ is at the third step of the profile). As the SLD of iron oxide is well above polymer values ($Nb_{Fe2O3}$=7×10$^{-6}$ Å$^{-2}$), we deduce that this local raise of SLD might evidence the confinement of the MNPs between the PS blocks of the first ($L_1$) and the second ($L_2$) bilayer (and between the second and the third for the thickest films with five bilayers). This irregular repartition of the MNPs not penetrating indistinctively all the space between PS layers might be related to a contribution of sedimentation during the self-assembly process. To complete this study on the insertion of MNPs into lamellae of PBMA-b-PS, AFM pictures of the doped film nb. 10 (with the highest load $\Phi$=0.15% of MNPs) and of the thicker film nb. 13 are shown on ESI-Figure 2. They exhibit a few differences with the AFM pictures of the undoped film (Figure 6). At first a much higher density of cracks are present compared to the surface of the pure BCP film. However this roughness remains of the order of a few nm at most (4.2±0.7 nm), otherwise the NR curves of doped films would not present oscillations. Another salient feature is the presence of dark disks on the phase image only, contrary to a pure BCP film showing dark disks – assimilated with holes – on both the height and the phase images. The contrast of phase AFM pictures being proportional to the surface toughness,[65] we can assign those dark regions to rubbery PBMA domains, much softer than the glassy PS surface. This observation is coherent with the NR and contact angles analyses, that both concluded that the top surface was mainly composed of a PS layer. On the opposite, the streaks on the height image appear on the phase image brighter than the PS background. This might be an indication of the presence of a hard layer of inorganic MNPs beneath the PS layer, somehow responsible for the appearance of cracks. A closer look to those streaks enabled to visualize the BCP lamellae directly, *i. e.* with an orientation of the layers parallel to the top surface (ESI-Figure 2). This phenomenon indicates some distortion of the layers induced by the MNPs, even though most of the surface area covered by the film remains in the same orientation as the pure copolymer.

*Refractive index enhancement and optical reflectivity*. In addition to the overall thickness of the films, the ellipsometric measurements provide their average refractive index independently. For a composite multi-layer nanostructure, RI is expected to exhibit a complex variation (*i.e.* varying with several parameters such as the total thickness of the coating, the number and the nature of the layers…). Nevertheless, we observe a monotonous behavior for the series of films at approximately constant thickness 110±4 nm and varying inorganic volume fraction $\Phi$ between 0.025% and 0.15% v/v. We see indeed that the values of $n$ increase linearly (1.56, 1.61, 1.64, 1.66) as a function of the average bilayer thickness $L_{ave}$ (39.3, 41.1, 42.2, 43.8 nm), which is not the case as a function of $\Phi$. Concerning the capability of those magnetic multi-lamellar films as photonic materials, we calculate the theoretical peak of their optical reflectivity[8] occurs at a wavelength $\lambda_{max}=4nL_{ave}\approx290$ nm for the maximum loading in MNPs (film nb. 10 on Table 6). If we intend to use our system in the visible instead of the ultra-violet range, we need to shift $\lambda_{max}$ to *e.g.* 500 nm. If such a factor around 1.7 was obtained only by the increase of the lamellar period, $L_{ave}$ should reach $\approx75$ nm. This would necessitate synthesizing a PBMA-*b*-PS sample of molar mass $M_n\approx270000$ g.mol[-1] according to the scaling law plotted on ESI-Figure 3. Although challenging, this goal seems possible to reach by ATRP. The doping by maghemite MNPs will favor the further decrease of this threshold $M_n$, because they were shown to increase both the lamellar period and the refractive index. By taking into account the observed empirical law that the refractive index of films doped with MNPs varies linearly with the average bilayer thickness, we expect $\lambda_{max}$ to scale with $L_{ave}^2$. Therefore a 30% increase of the period would be sufficient to reach a maximum $\lambda_{max}\approx500$ nm in the central part of the visible spectrum. Such a bilayer thickness $L_{ave}\approx57$ nm could be reached for a molar mass $M_n\approx190000$ g.mol[-1], as estimated from the master curve plotted from the bibliographic study.

*5. Conclusions*

A well-defined nearly symmetrical PBMA-*b*-PS diblock copolymer with a number-average molar mass as high as 112 000 g.mol[-1] and a polydispersity index as low as 1.4 has been successfully synthesized by controlled radical polymerization *via* ATRP. At first, the lamellar ordering of thin films formed with this diblock copolymer after annealing has been evidenced by AFM by the presence of "islands" and "holes" type's defects on top of the films.

Our samples differ from the previously reported systems in two main points: the polymers were synthesized by atom transfer radical polymerization rather than anionic polymerization and the nanoparticles were coated by a "grafting from" method rather than "grafting onto". These differences of the chemical route modifying the end-groups of the polymer chains are invoked to explain an order inversion of the repetition sequences in the multi-lamellar films on flat silicon substrates, as revealed by neutron reflectivity and checked by contact angle measurements and AFM phase analysis. In other words, the chemical end-groups of the copolymer have a strong influence on the surface energies of the lamellae at the interfaces with both substrate and air.

In parallel, magnetic iron oxide nanoparticles with a coating of PS chains were prepared. As for the copolymer, these chains were synthesized by ATRP, with the supplemental difficulty of initiating the polymerization at a nanometric surface. Their molar mass dispersity was decreased down to 1.4 thank to the choice of magnetic cores with a narrower distribution of diameters. The presence of clusters presumably due to coupling reactions between growing chains from adjacent nanoparticles could not be totally eliminated. A recent study showed that the clustering effect of MNPs also occurs in a homogeneous matrix of PS,[66] even though they were previously properly dispersed in solution. If the average size of the clusters remains below a threshold value of the order of the lamella thickness (*e.g.* 50 nm) and if they are not too dense, they can be flattened between the lamellae and remain compatible with the self-assembly process. The TEM images of Figure 4 that are projections on the flat surface of the copper grid show indeed aggregates that do not seem thicker that one layer of MNPs.

The introduction of PS-grafted γ-Fe$_2$O$_3$ magnetic nanoparticles into the thin films leads to nanocomposites with a preserved lamellar structure, opening possibilities for applications as reflectors in different spectral ranges. In particular, ellipsometry shows a noticeable increase of the optical refractive index that encourages studying further those multi-lamellar magnetic films with a polymer matrix as a material for optical waveguides. Our long-term prospect is to take benefit from the orientation

This document is the author manuscript version of an article that appeared in final form in *Polymer*, 2010, *51*, 4673–4685, after peer review and technical editing by the publisher. DOI: 10.1016/j.polymer.2010.08.043properties under magnetic field of such nanocomposite thin films for technical applications as responsive mirrors, specific wave absorption or reflection media in either the visible or hyper-frequencies spectrum. In particular, we plan to study the optical reflectivity of these magnetic nanocomposite films under the application of a constant magnetic field of the order of 1 Tesla: the induced magnetization of MNPs confined in 2-d layers associated with the rubbery behavior of PBMA in the "sandwich" nanostructure could lead to a modulation of the lamellar period through a magneto-striction mechanism.[67] Such an effect would lead to photonic materials that could be controlled by a magnetic field. Other applications related to electronics industry could be thought of, the prerequisite of keeping temperature below *e.g.* 450°C during the whole nano-structuration process to be integrated with semi-conductors being fulfilled for the types of BCPs and MNPs that we chose. Those applications necessitate further physical studies (*e. g.* microwave absorption measurements) of the pure lamellar BCP films and of the samples doped with MNPs.

*Acknowledgements*
- Fabrice Cousin and Frédéric Ott for their help with neutron reflectivity on the EROS setup at the LLB CEA-Saclay facility;
- Yvette Tran from the "Laboratory de Physico-Chimie des Polymères et des Milieux Dispersés" at ESPCI ParisTech for access to ellipsometry;
- Sanae Tabnaoui, Romain Pintat and Maeva Serror for their participation to the work during internships;
- Fabrice Audouin from the LCP laboratory at UPMC for his practical help with SEC;
- Emanuel Lepleux (Scientec, Les Ulis, France) for a demonstration experiment with the PicoScan AFM from Molecular Imaging;
- Serge Durand-Vidal for access to the DI Nanoscope III AFM setup of the PECSA laboratory in Paris and Emmanuel Ibarboure for his expertise to interpret AFM measurements and for complementary images provided in the ESI file taken with the instrument of LCPO in Pessac;
- One of us (S. D-M) is grateful to Pr. Jean-Jacques Rousseau from University Jean-Monnet and to Pr. Jean Lemerle from SIFAC association for partial funding of her fellowship.

*Appendix. Supplementary data*
Supplementary data associated with this article can be found online at DOI: 10.1016/j.polymer.2010.08.043.

**ESI-Figure 1.** Magnetization curves of "polydisperse and "monodisperse" $\gamma$-Fe$_2$O$_3$ magnetic nanoparticles, before and after coating by PS chains.

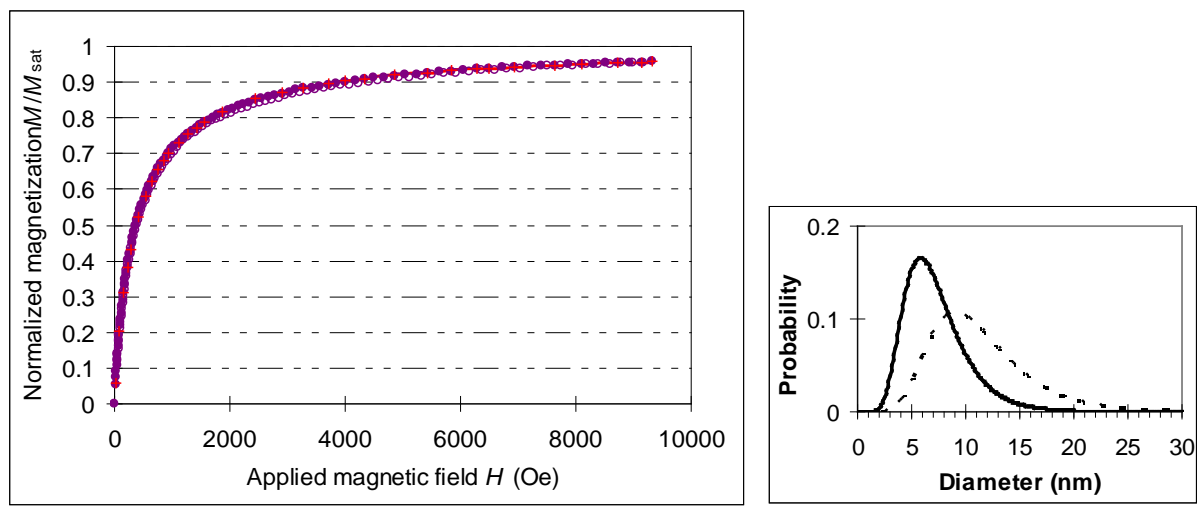

**(a)** "Polydisperse" sample of MNPs right after synthesis in water (HNO$_3$ pH≈1.7). The curve is fitted by Langevin's equation of superparamagnetism weighted by a Log-normal distribution diameters of parameters $d_0$=6.8 nm and $\sigma$=0.39 plotted in number (solid line) and volume probability (dashed line). The saturation plateau writes $M_{sat}=m_{spe}\Phi$ where the volume fraction is $\Phi$=1.68% and the specific magnetization $m_s$ = 3.1×10$^5$ A/m. Expressed as a number of Bohr's magnetons, the average magnetic dipole of the suspension taking into account the width of the size distribution is around 4×10$^4$ $\mu_B$, which is also approximately the number of iron+III ions per MNP. The curve is perfectly reversible for increasing (filled symbols) and decreasing field values (empty ones).

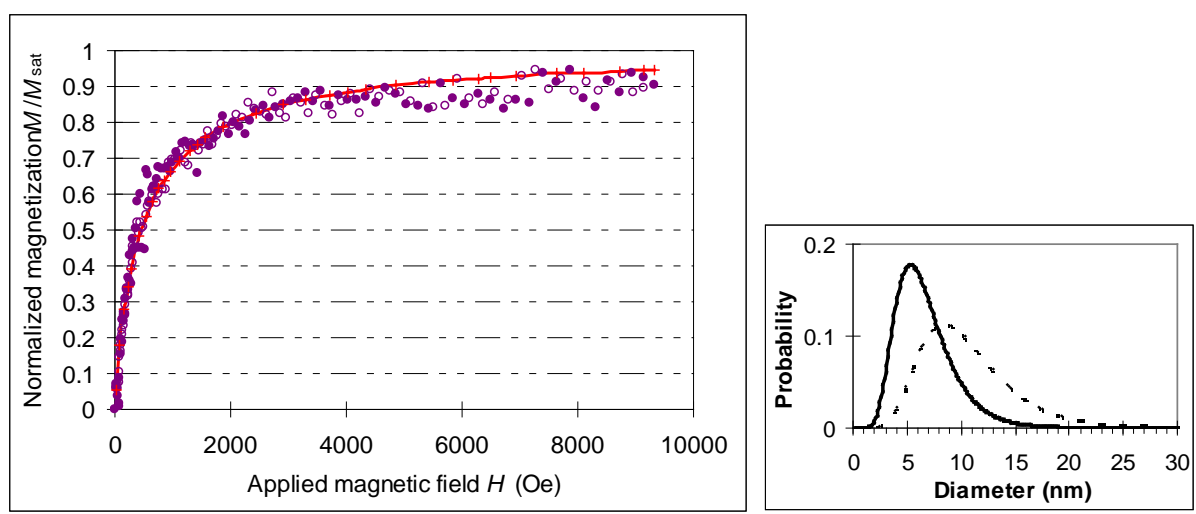

**(b)** "Polydisperse" sample $\gamma$-Fe$_2$O$_3$@PS1 after coating by PS chains and dispersion in dichloromethane. The curve is fitted with the same parameters as (a) except $\Phi$=0.0038%.

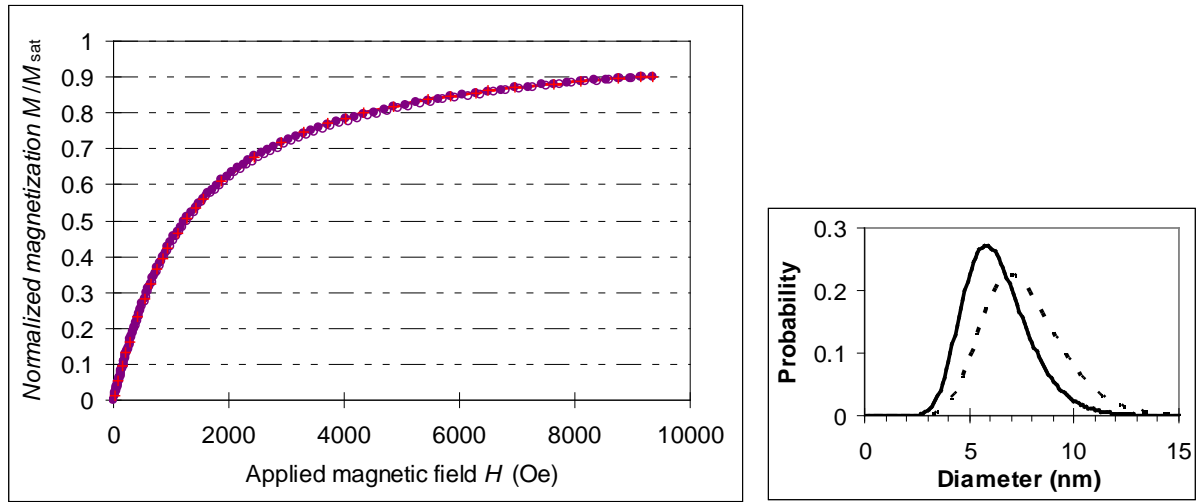

**(c)** "Monodisperse" sample of MNPs obtained by successive phase separations with excess $HNO_3$ after coating by oleic acid and dispersion in n-hexane. The curve is fitted by Langevin's equation of superparamagnetism weighted by a Log-normal distribution diameters of parameters $d_0$=6.2 nm and $\sigma$=0.25 plotted in number (solid line) and volume probability (dashed line). The saturation plateau writes $M_{sat}=m_{spe}\Phi$ where the volume fraction is $\Phi$=0.66% and the specific magnetization $m_s$= 2.7×10⁵ A/m. The average magnetic dipole of the suspension is around 8×10³ $\mu_B$, which is significantly lower than the previous case due to the narrower size distribution. The curve is perfectly reversible for increasing (filled symbols) and decreasing field values (empty ones).

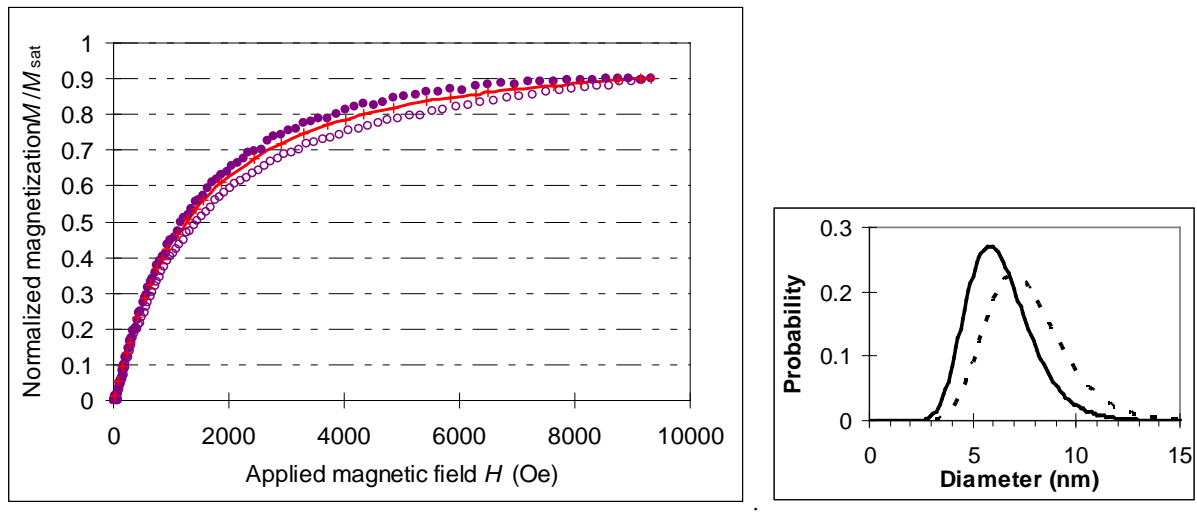

**(d)** "Monodisperse" sample $\gamma$-$Fe_2O_3$@PS2 after coating by PS chains and dispersion in dichloromethane. The curve is fitted with the same parameters as (c) except $\Phi$=0.018%. The imperfect reversibility of the curve is ascribed to a slight aggregation of the MNPs.

*Appendix. Supplementary data to article appeared in Polymer 2010, 51, 4673–4685,* [DOI: 10.1016/j.polymer.2010.08.043](DOI: 10.1016/j.polymer.2010.08.043)

**ESI-Figure 2.** Geometrical analysis of defects at the surface of the undoped $PBMA_{425}$-*b*-$PS_{490}$ film (nb. 2) and of doped ones (nb. 10 and 13) measured from AFM height, amplitude and phase images taken with either a DI Nanoscope or a MI Picoscan microscope.

- <u>Analysis of Figure 6(a)</u> taken with a Nanoscope III microscope in the dry Tapping Mode™ showing islands-type defects. The solid line fitting the statistics of heights over the whole picture is a Gaussian law with mean value $H_0$=16.37 nm and standard deviation $\sigma$=0.73 nm.

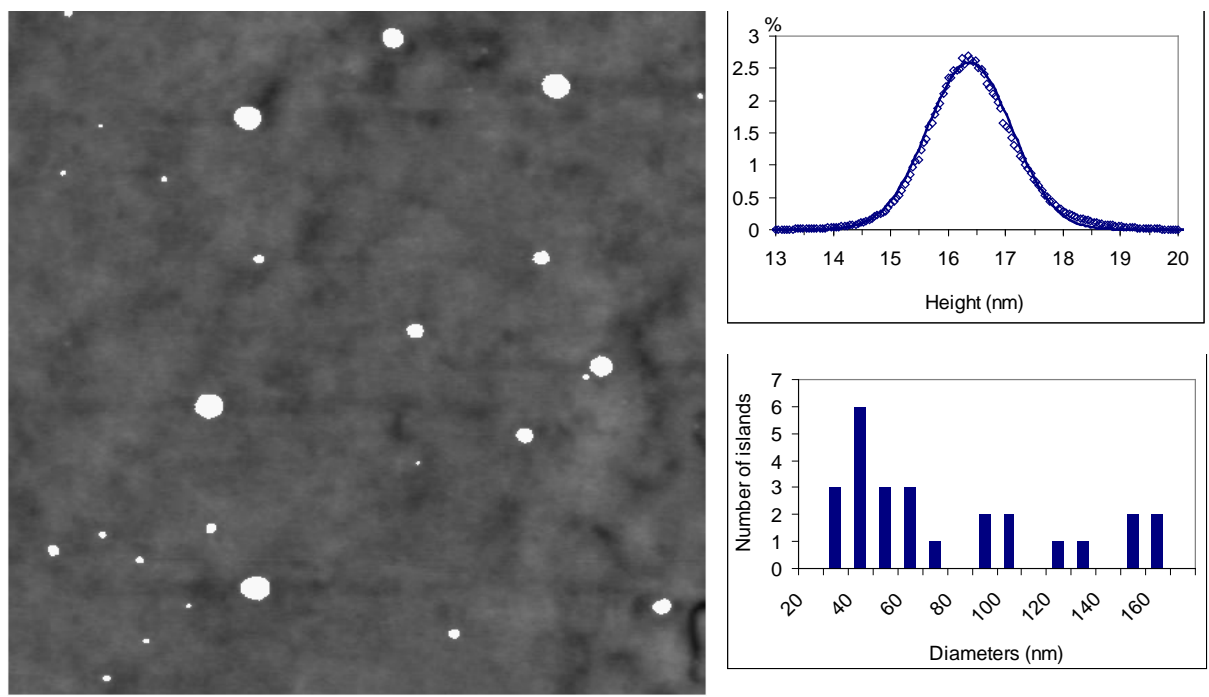

- <u>Other topographical pictures</u> showing loops scratched at the surface of the same film (nb. 2).

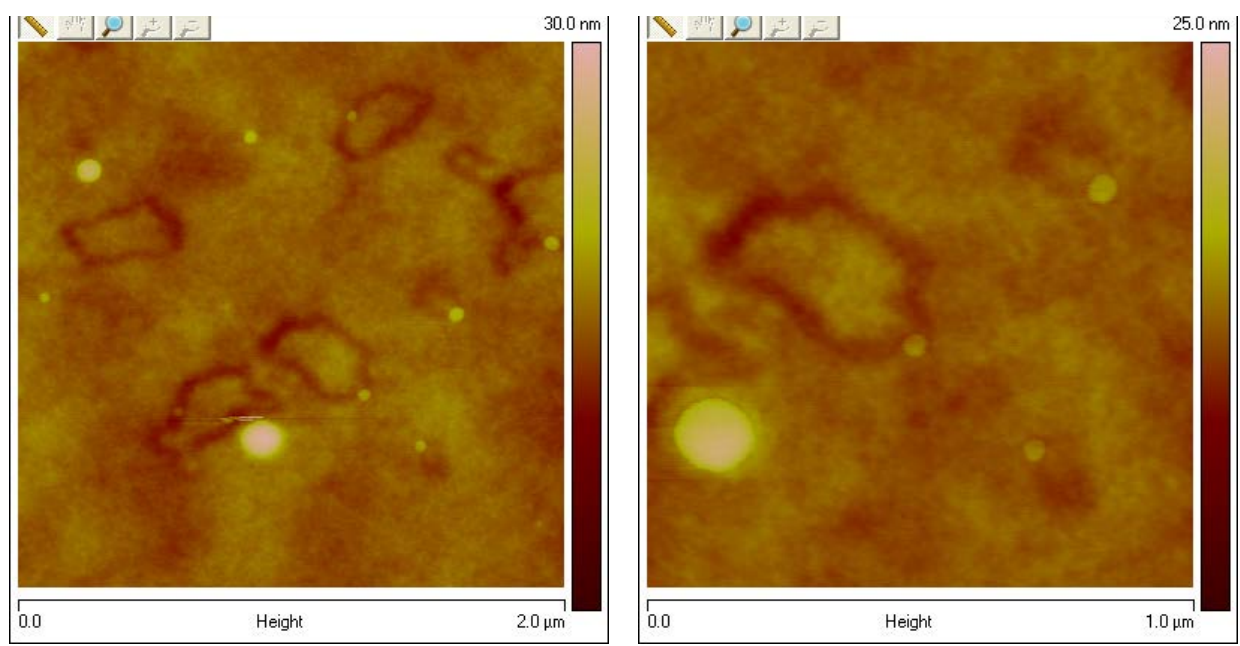

- <u>Zoom around a particularly large island</u> with vertical cross-sections indicated by the crosses.

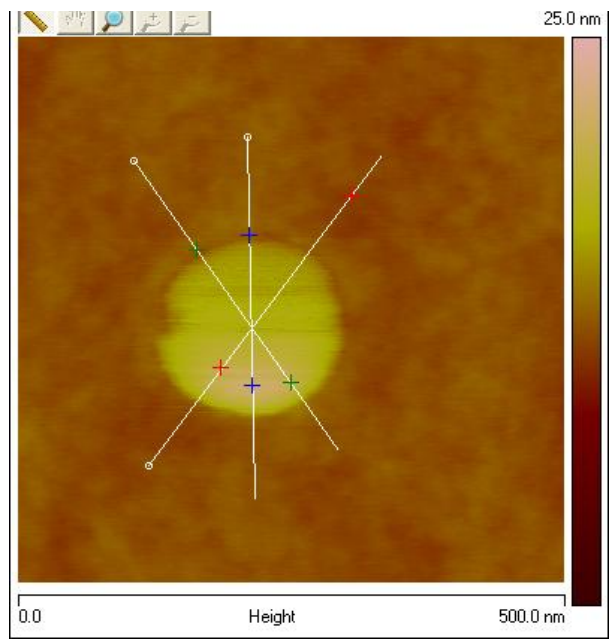
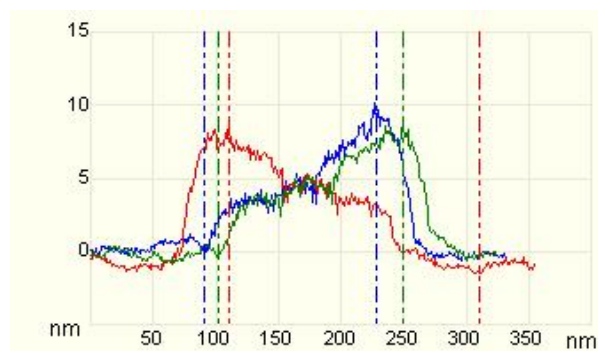

| Crosses | Horizontal Distance | Vertical Distance | Surface Distance | Theta |
|---|---|---|---|---|
| Blue | 137.994 (nm) | 9.179 (nm) | 155.023 (nm) | 3.805 (º) |
| Red | 199.364 (nm) | -9.649 (nm) | 210.864 (nm) | -2.771 (º) |
| Green | 148.513 (nm) | 8.725 (nm) | 159.128 (nm) | 3.362 (º) |

- <u>Measurement of the depths of holes</u> at the surface of the undoped film nb.2

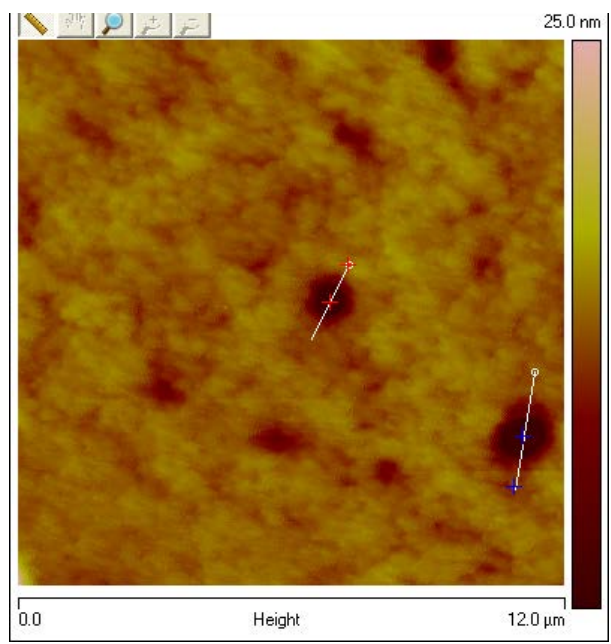
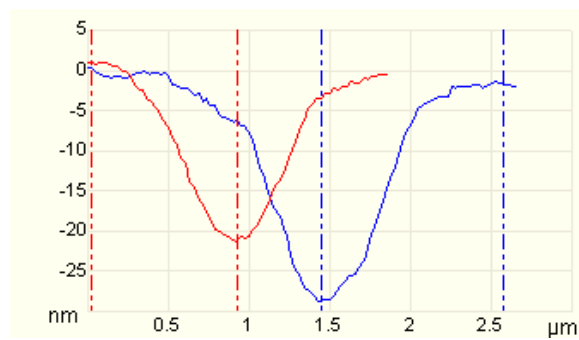

| Crosses | Horizontal Distance | Vertical Distance | Surface Distance | Theta |
|---|---|---|---|---|
| Blue | 1.118 (µm) | 27.015 (nm) | 1.119 (µm) | 1.384 (º) |
| Red | 0.895 (µm) | -22.148 (nm) | 0.896 (µm) | -1.417 (º) |

- Analysis of Figure 6(b) taken on film 2 by a Picoscan microscope in Acoustic AC Mode$^{TM}$.

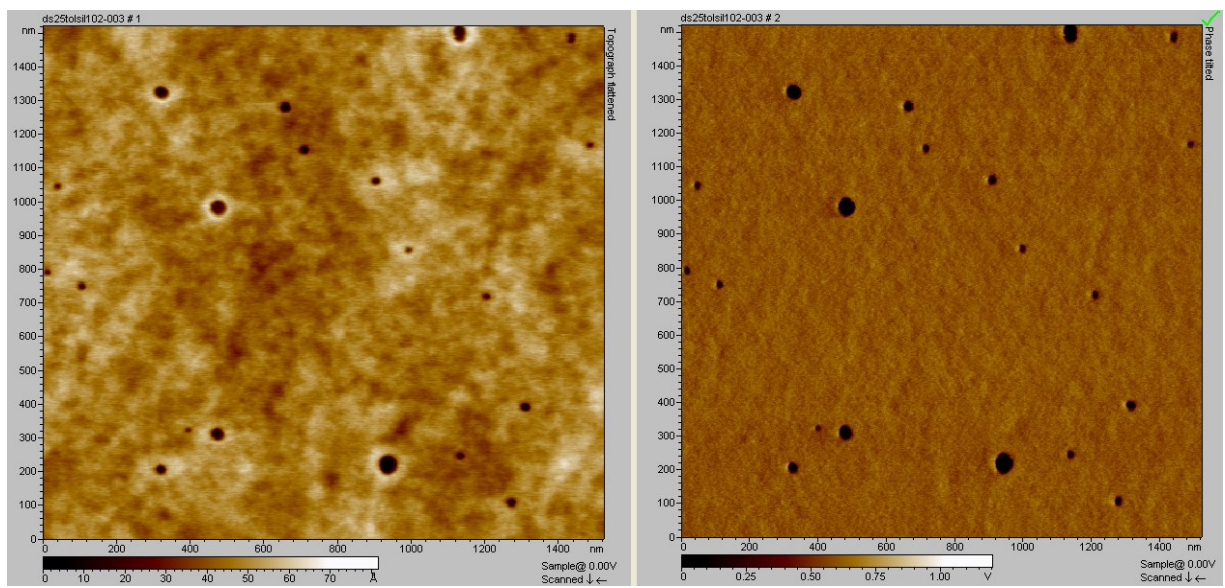

Height image (1.52µm×1.52µm)　　　　　　　　　Phase image (1.52µm×1.52µm)

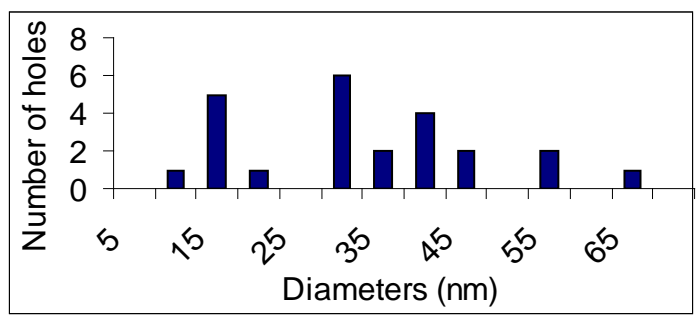

- Optical micrographs of an undoped film (nb. 2) and a doped one (nb. 10) under the AFM tip.

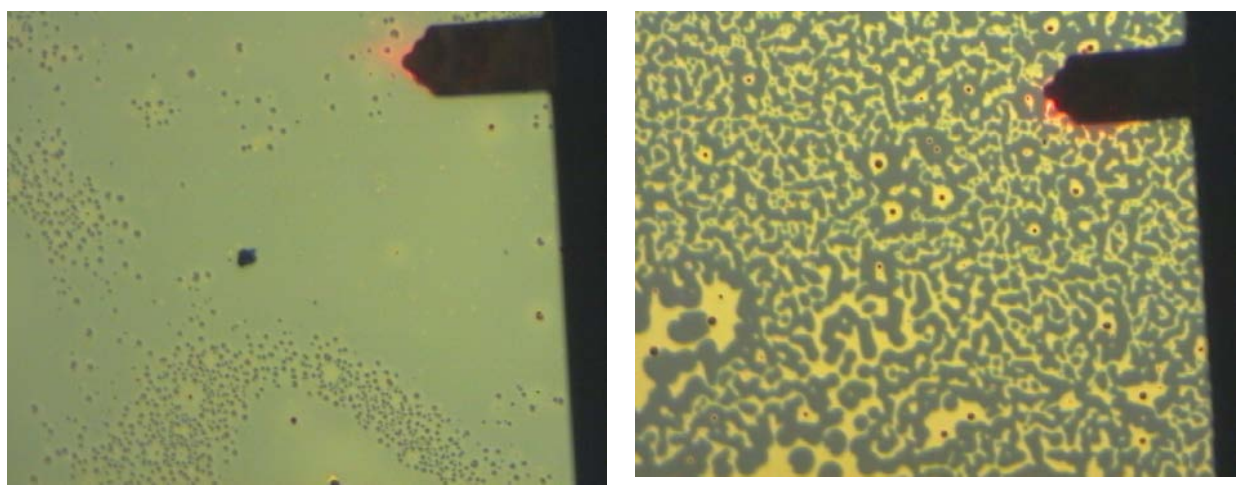

*Appendix. Supplementary data to article appeared in Polymer 2010, 51, 4673–4685, DOI: 10.1016/j.polymer.2010.08.043*

- <u>Scratches at the surface of a doped film at the maximum loading in MNPs</u> (nb. 10) measured on AFM images taken with a Nanoscope III microscope in the Tapping Mode™.

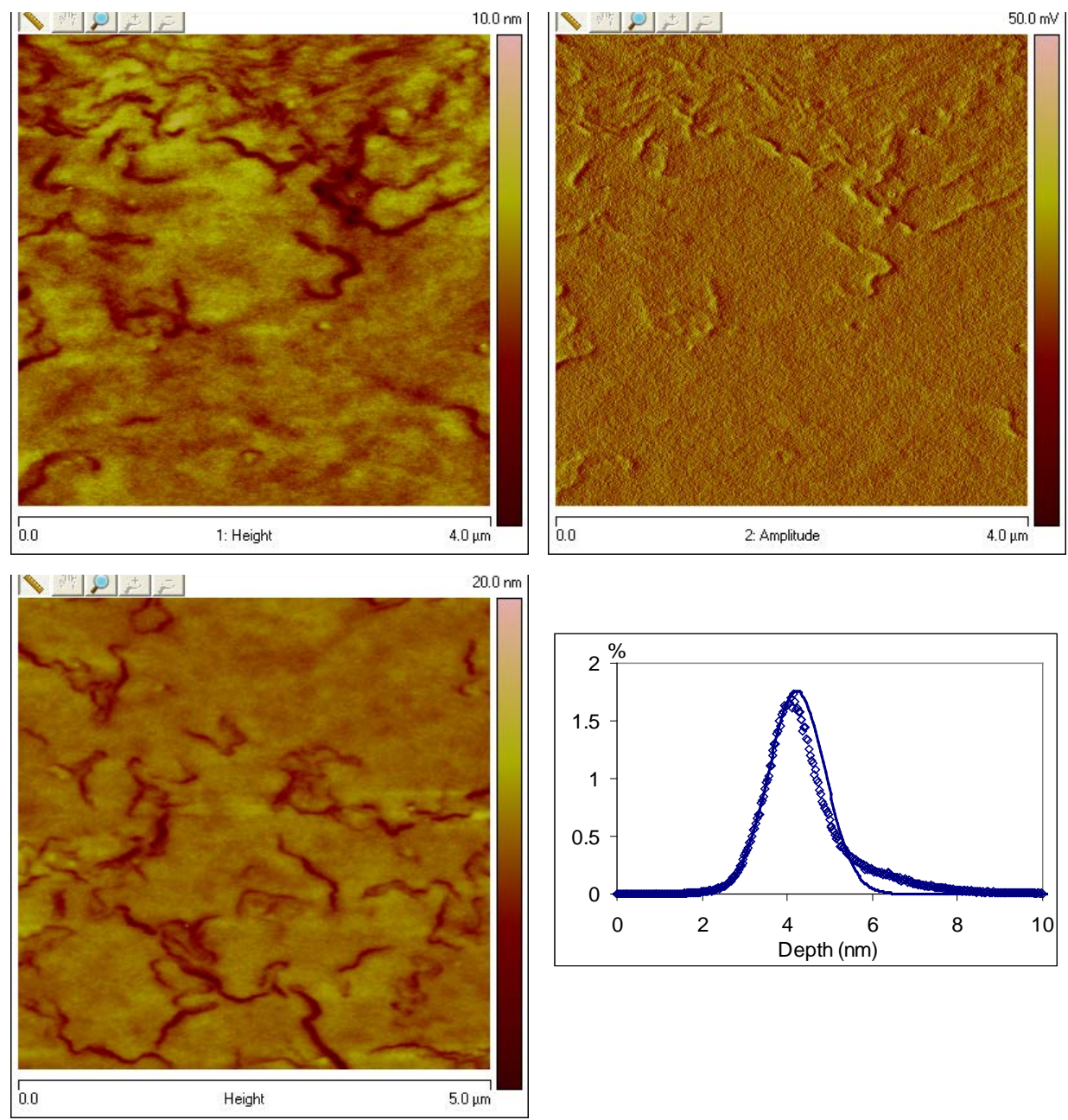

The histogram of the depths of the cracks at the surface of the doped film (nb. 10, with maximum load $\Phi$=0.15% of MNPs) can be approximated by a Gaussian law 4.23±0.66 nm.

- <u>Analysis of the defects at the surface of the doped film nb. 13:</u> $PBMA_{425}$-*b*-$PS_{490}$ copolymer doped with $\gamma$-$Fe_2O_3$@PS1 nanoparticles at $\Phi$ = 0.025% and annealed 72h at 150°C.

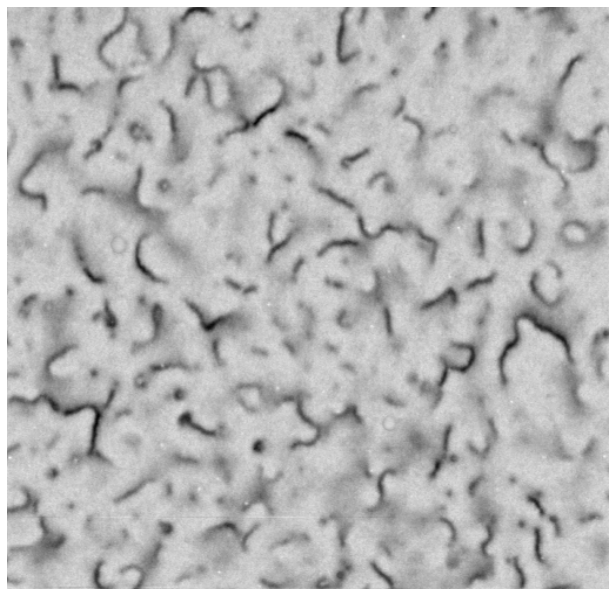
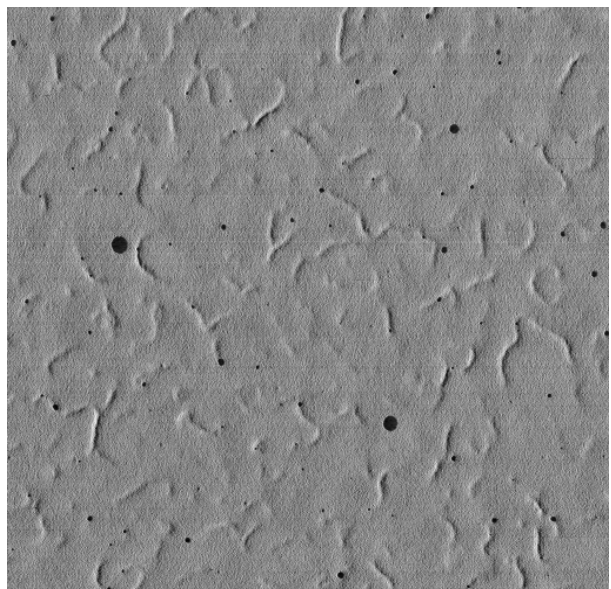

Picoscan™     Height image          5μm×5μm          Phase image

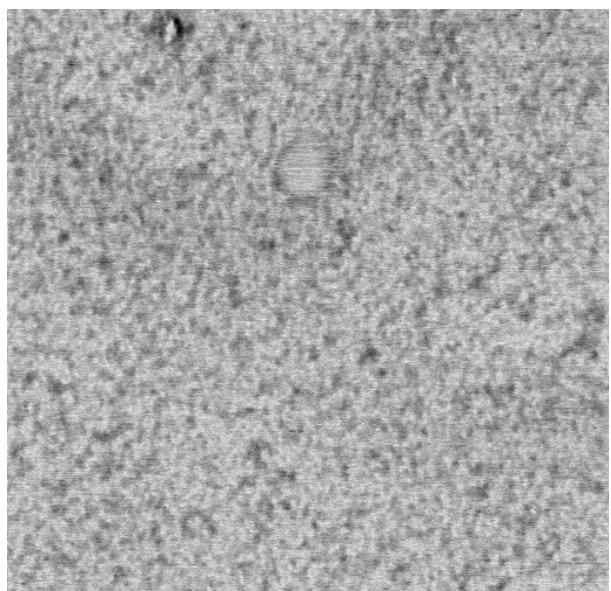
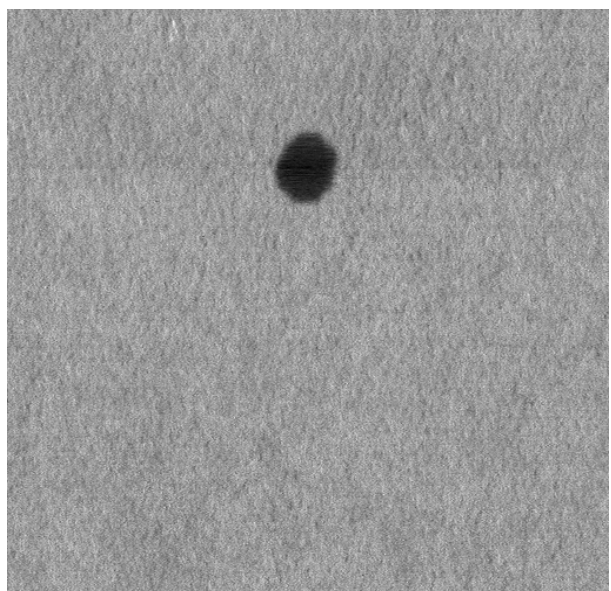

Picoscan™     Height image          413nm×413nm      Phase image

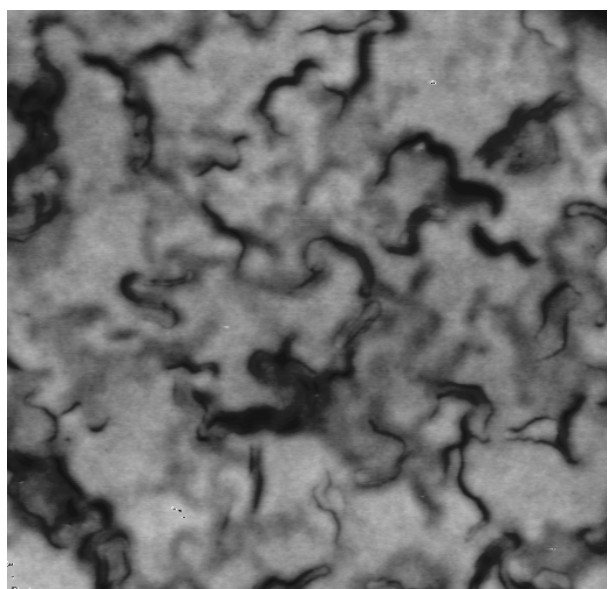
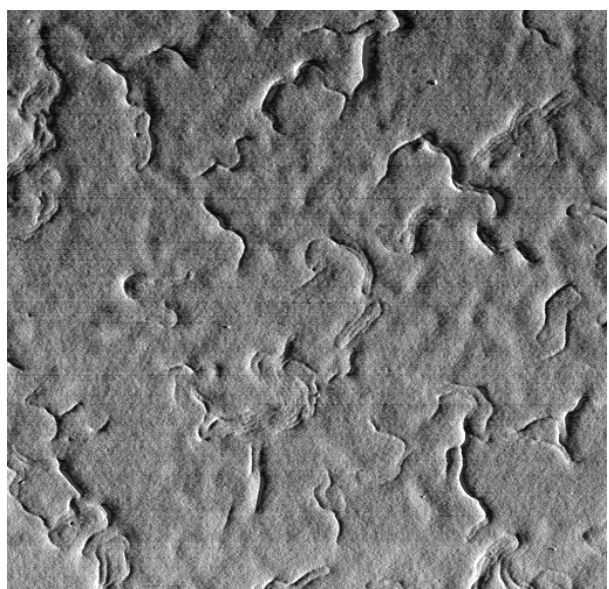

Nanoscope™    Height image          10μm×10μm        Phase image

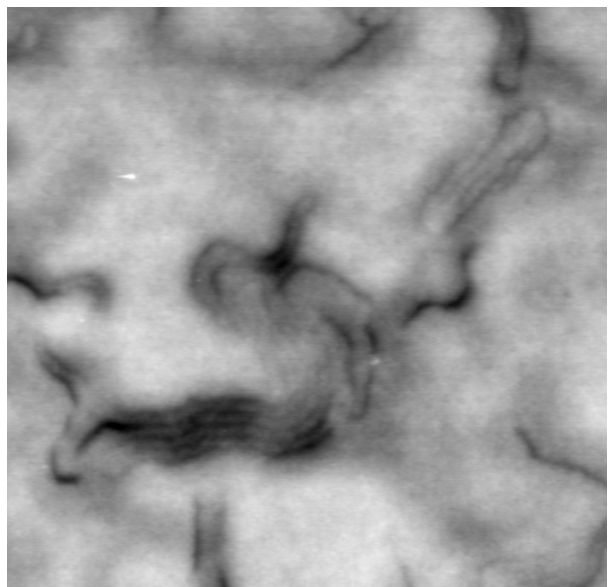
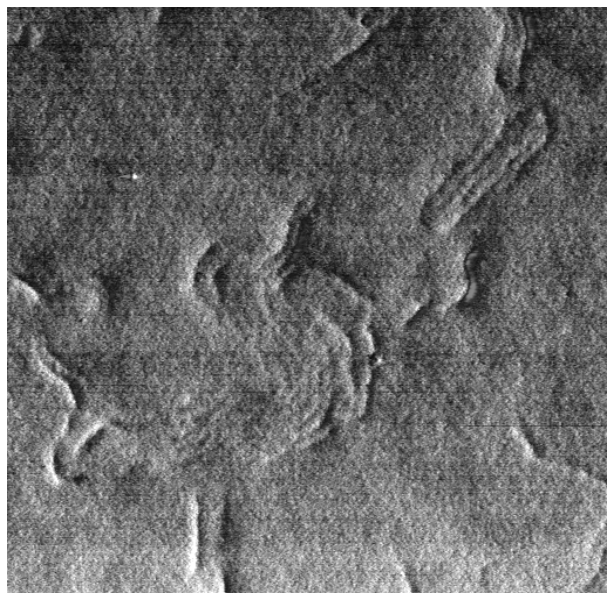

Nanoscope™        Height image        3.7μm×3.7μm        Phase image

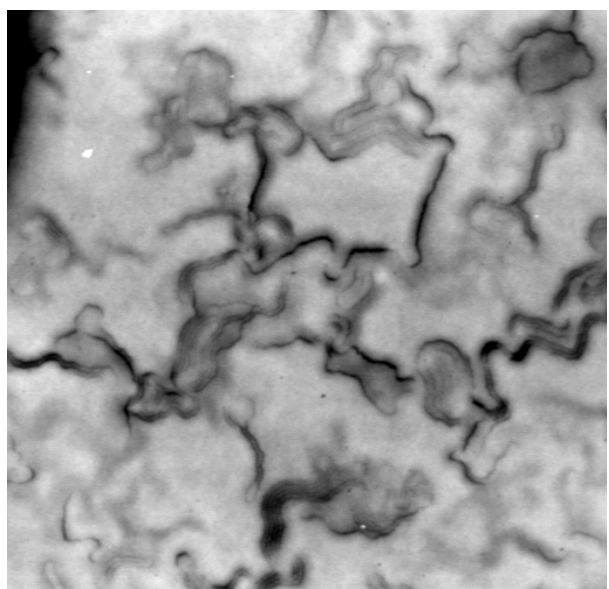
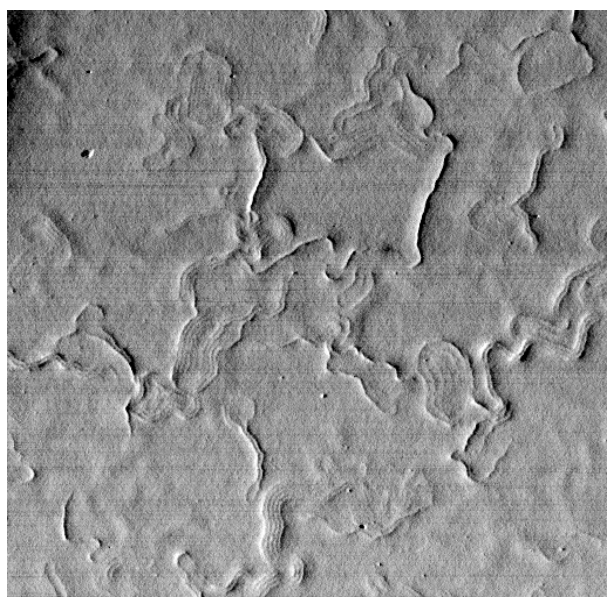

Nanoscope™        Height image        10μm×10μm        Phase image

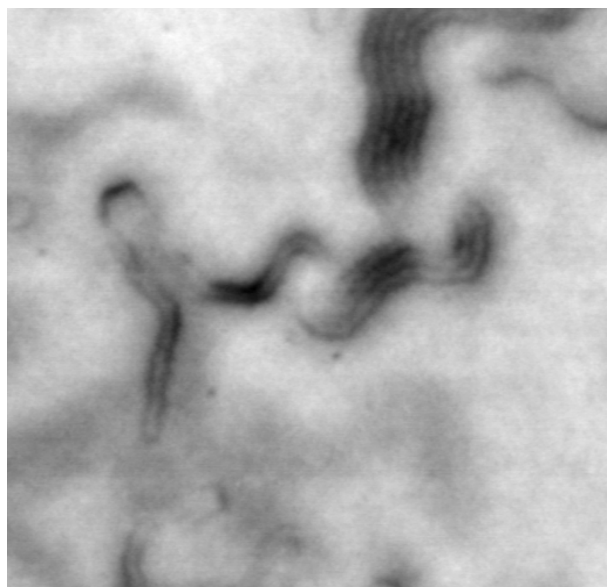
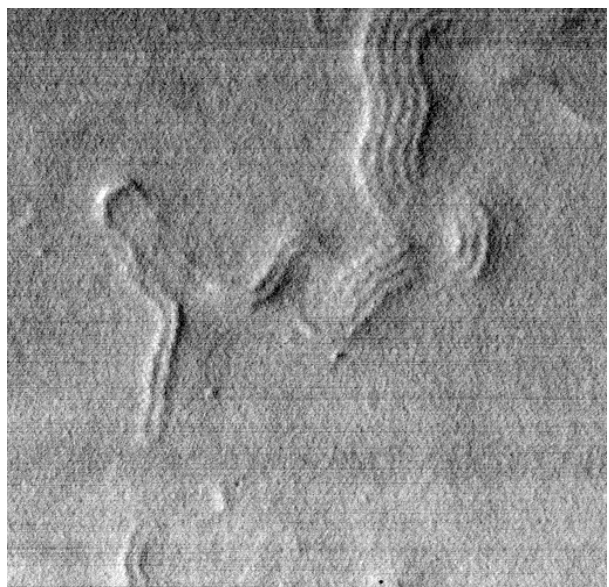

Nanoscope™        Height image        3.3μm×3.3μm        Phase image

*Appendix. Supplementary data to article appeared in Polymer 2010, 51, 4673–4685,* DOI: 10.1016/j.polymer.2010.08.043

**ESI-Figure 3.** Bibliographic study of the scaling law between the lamellar period $L_{ave}$ and the molar mass Mn. Triangles: measured by Small Angle X-ray Scattering for PS-b-PI 1; Circles: measured by SAXS for PS-b-PBMA 2 or PS-b-PBMA-MAH 3 and SANS for dPS-b-PBMA 4; Crosses: measured by Neutron Reflectivity for (d)PS-b-PBMA 5, 6 or PBMA-b-PS (this work); Asterisks: measured from the height of the film defects at the top surface either by optical interference microscopy7 or AFM6, 8 for (d)PS-b-PBMA or PBMA-b-PS (this work); Squares: measured by X-Ray Reflectivity for (d)PS-b-PBMA.6, 9 The polymers were always annealed in the range 150°C-165°C, except for molar masses below 70000 g.mol-1 for which they were necessarily heated up between 200°C and 260°C to induce the lamellar ordering.

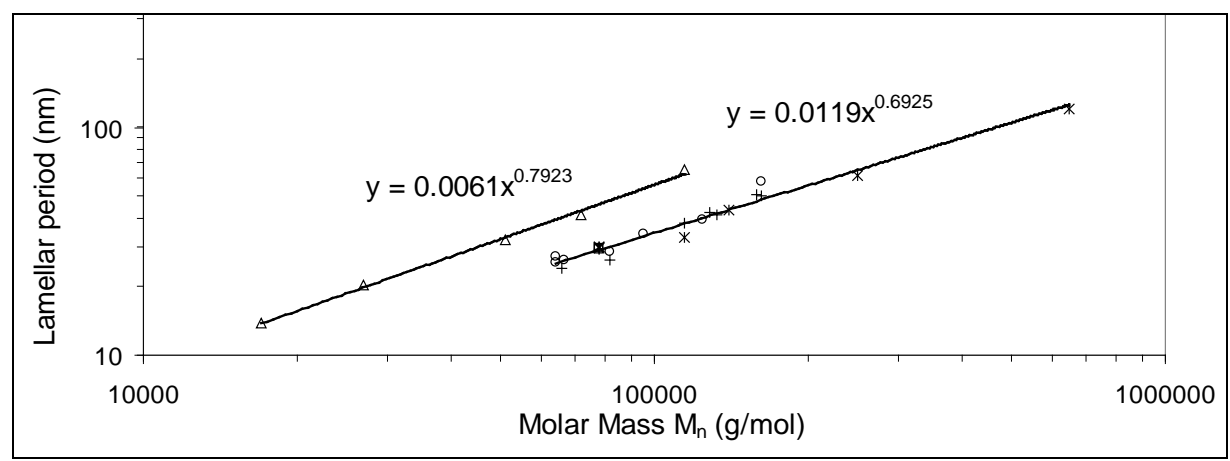

**ESI-Scheme 1.** Structure of the **(A)** PS-*b*-PBMA diblock copolymer synthesized by anionic polymerization for structural analysis reported in references 9–16, 19–24 of main text and of the **(B)** PBMA-*b*-PS diblock copolymer synthesized by ATRP as described in part 2.2.

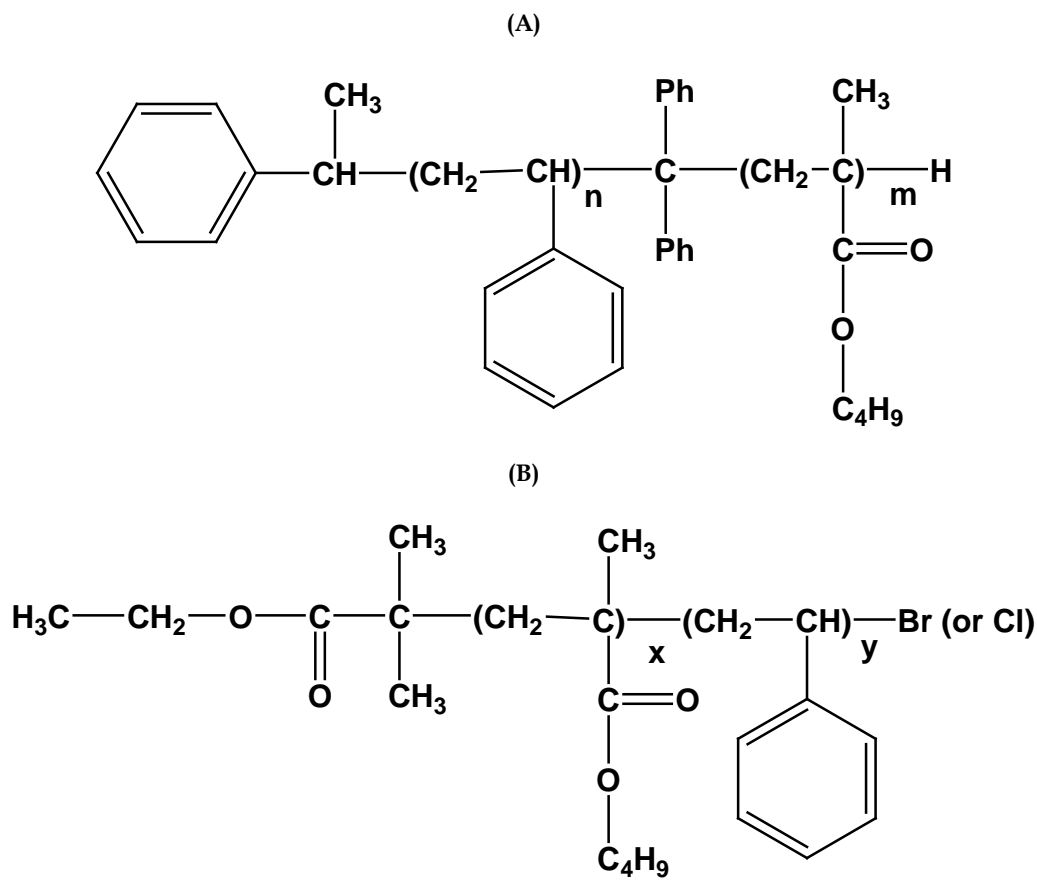